\documentclass[12pt]{article}
\usepackage{amsfonts,amsmath,amssymb,epsf,framed}
\usepackage{amsmath,braket}
\usepackage{graphicx,hyperref,color,comment}
\usepackage{cite}

\usepackage{subfigure}

\edef\restoreparindent{\parindent=\the\parindent\relax}
\usepackage{parskip}
\restoreparindent

\hypersetup{
    bookmarks=true,         
    unicode=false,          
    pdftoolbar=true,        
    pdfmenubar=true,        
    linktocpage=true,       
    pdffitwindow=false,     
    pdfstartview={FitH},    
    pdfnewwindow=true,      
    colorlinks=true,       
    linkcolor=blue,          
    citecolor=blue,        
    filecolor=blue,      
    urlcolor=blue           
}
\numberwithin{equation}{section}									

\DeclareMathOperator{\tr}{tr}

\newcommand{\onshell}{\scriptsize\mbox{on-shell}}
\let\a=\alpha \let\b=\beta \let\c=\chi \let\d=\delta \let\e=\epsilon \let\ve=\varepsilon \let\g=\gamma    \let\m=\mu \let\n=\nu
 \let\p=\phi \let\r=\rho \let\s=\sigma \let\t=\tau \let\th=\theta  \let\vp=\varphi   
   \let\L=\Lambda \let\O=\Omega \let\P=\Phi     
\def\nn{\nonumber}
\def\inf{\infty}

\def\pa{\partial}

\def\wtd{\widetilde}

\DeclareMathOperator{\Tr}{Tr}

\begin{document}

\begin{titlepage}
\thispagestyle{empty}

\vspace*{-2cm}
\begin{flushright}
RUP-23-16\\
YITP-23-114\\
\vspace{1cm}
\end{flushright}

\bigskip

\begin{center}
\noindent{{\Large \textbf{Dimensional Reduction of the $S^3$/WZW Duality}}}\\
\vspace{2cm}

Kenta Suzuki$^a$ and \ Yusuke Taki$^b$
\vspace{1cm}\\

{\it $^a$ Department of Physics, Rikkyo University, Toshima, Tokyo 171-8501, Japan}\\[6pt]
{\it $^b$ Center for Gravitational Physics and Quantum Information,\\
Yukawa Institute for Theoretical Physics,\\
Kyoto University, Sakyo-ku, Kyoto 606-8502, Japan}\\
\vspace{1mm}

\bigskip \bigskip
\vskip 3em
\end{center}

\begin{abstract}
Recently proposed duality relates the critical level limit $\hat{k} \to -2$ of $SU(2)_{\hat{k}}$ WZW models to a classical three-dimensional Einstein gravity on a sphere.
In this paper, we propose a dimensional reduced version of this duality.
The gravity side is reduced to a Jackiw-Teitelboim (JT) gravity on $S^2$ with a non-standard boundary term, or a BF theory with $SU(2)$ gauge symmetry.
At least in low temperature limit, these two-dimensional gravity theories completely capture the original three-dimensional gravity effect.
The CFT side is reduced to a certain complex Liouville quantum mechanics (LQM) with $SU(2)$ gauge symmetry.
Our proposal gives an interesting example of a holography without boundary.
We also discuss a higher-spin generalization with $SU(N)$ gauge symmetry.

\end{abstract}

\end{titlepage}

\newpage

\tableofcontents

\section{Introduction}
\label{sec:introduction}
The AdS/CFT correspondence \cite{Maldacena:1997re,Gubser:1998bc,Witten:1998qj} has provided remarkable understanding of quantum gravity theories,
including insights toward resolving the black hole information paradox, and strongly interacting quantum field theories.
This correspondence relates quantum gravity theories on $d + 1$ dimensional asymptotically anti-de Sitter spaces (AdS$_{d+1}$)
to certain class of $d$ dimensional strongly interacting conformal field theories (CFT$_d$) with a large amount of degrees of freedom.
Beyond asymptotically AdS spacetime, however, our understanding of quantum gravity is highly limited,
even though there have been several proposals for gauge/gravity correspondences in non-AdS spacetimes.
Some partial examples of these proposals are the dS/CFT correspondence \cite{Strominger:2001pn, Maldacena:2002vr,Witten:2001kn} for asymptotically de Sitter (dS) spacetime,
the celestial holography \cite{Pasterski:2016qvg, Raclariu:2021zjz, Pasterski:2021rjz} for asymptotically Minkowski spacetime,
and the holography without boundaries \cite{Kawamoto:2023wzj}.

In particular, the papers \cite{Hikida:2021ese, Hikida:2022ltr} proposed a holographic duality between a classical gravity on $S^3$
and $SU(2)_{\hat{k}}$ Wess-Zumino-Witten (WZW) model with a certain analytically continued value of the level $\hat{k}$.
We call this duality the $S^3$/WZW duality in the rest of the paper.
Even though their original motivation of this proposal was to study Lorentzian de Sitter spacetime,
this duality provides an interesting explicit model for the holography without boundaries.
They used this Euclidean version of duality to prepare the Hartle-Hawking wave functional \cite{Hartle:1983ai} for the Lorentzian three-dimensional de Sitter spacetime. This duality has some ambiguity at the subleading order of the Newton constant (i.e. one-loop level),
but the leading classical correspondence was explicitly checked in \cite{Hikida:2021ese, Hikida:2022ltr}.
Therefore, in this paper, we focus on this Euclidean version of the duality in the leading classical order.

In this paper, we first note that the leading classical on-shell action of the theories considered in \cite{Hikida:2021ese, Hikida:2022ltr} are just linear in temperature,
so the $S^3$/WZW duality is completely captured by a low-temperature dynamics.
This reminds us about the story of the near-AdS$_2$/near-CFT$_1$ correspondence \cite{Almheiri:2014cka, Jensen:2016pah, Maldacena:2016upp, Engelsoy:2016xyb}
which universally appears in the near-extremal limit of higher dimensional black holes \cite{Ghosh:2019rcj}.
A wide class of near-extremal higher dimensional black holes develops an AdS$_2$ near horizon geometry,
and their low temperature dynamics is dominated by zero modes in the remaining extra dimensions.
Hence the $S$-wave reduction for the extra dimensions leads to the Jackiw-Teitelboim (JT) gravity \cite{Jackiw:1984je, Teitelboim:1983ux} in near-AdS$_2$ background.
The breaking of the global AdS$_2$ isometry (i.e. $SL(2, \mathbb{R})$) comes from the fluctuations of the boundary of AdS$_2$,
and the effect of this breaking is described by the so-called Schwarzian theory \cite{Maldacena:2016upp, Maldacena:2016hyu, Mertens:2017mtv, Mertens:2018fds}.
In this paper, we will show that a similar dimensional reduction is possible for the $S^3$/WZW duality considered in \cite{Hikida:2021ese, Hikida:2022ltr}.

The rest of the paper is organized as follows. In the remaining of this section, we will give a brief review of the $S^3$/WZW duality.
In section~\ref{sec:3d gravity}, we will discuss the dimensional reduction of the gravity side within the second order formalism.
This leads to the JT gravity but with a non-standard boundary term. 
Then, we move onto the first-order formalism of the three-dimensional gravity in terms of the Chern-Simons theory in section~\ref{sec:chern-simons}.
We clarify the boundary condition of this theory.
The dimensional reduction can also be discussed in the first-order formalism, which leads to the BF-theory with $SU(2)$ gauge symmetry.
We will explain this in section~\ref{sec:bf}, together with the higher spin $SU(N)$ generalizations.
In section~\ref{sec:wzw}, we move onto the dual field theory. We first review the WZW model and its reduction to a complex Alekseev-Shatashvili theory.
In section~\ref{sec:liouville-qm}, we finally discuss the dimensional reduction of the WZW model or the Alekseev-Shatashvili theory, which leads to a complex Liouville quantum mechanics (LQM).
This establishes our proposed duality between the classical limit of the JT gravity discussed in section~\ref{sec:3d gravity} and the particular limit of the level $\hat{k} \to -2$ of this complex LQM.
In the appendices, we summarize our notation for the $SU(2)$ and some supplemental materials.

\subsection{Review of the $S^3$/WZW duality}
As described above, the duality considered in this paper is based on \cite{Hikida:2021ese,Hikida:2022ltr}. In this subsection we briefly review the basics of the dS/CFT correspondence, the model used in \cite{Hikida:2021ese,Hikida:2022ltr}, and its relation to our Euclidean version of the duality. 

The dS/CFT correspondence \cite{Strominger:2001pn,Maldacena:2002vr,Witten:2001kn} (see also \cite{Maldacena:1998ih,Park:1998qk,Park:1998yw}) is a holographic duality for the (Lorentzian) de Sitter (dS) spacetime analogous to the AdS/CFT correspondence; the dual CFT$_d$ is thought of as living at the future asymptotic boundary $\mathcal{I}^+$ of dS$_{d+1}$. 
An analogous dictionary to GKPW relation \cite{Gubser:1998bc,Witten:1998qj} is formulated for the wave functional of universe \cite{Maldacena:2002vr}
\begin{align}
    \Psi_{\text{dS}}[\phi^{(0)}]=\left\langle \exp\left(\int d^dx\, \mathcal{O}(x)\phi^{(0)}(x)\right)\right\rangle_{\text{CFT}},
\end{align}
where the wave functional of universe $\Psi_{\text{dS}}[\phi^{(0)}]$ is defined as the path integral over all the bulk fields $\phi$ including the metric $g_{\mu\nu}$ with boundary conditions $\phi(x,T)|_{\mathcal{I}^+}=\phi^{(0)}(x)$, where $T$ denotes the global time of dS$_{d+1}$.
In the CFT side, $\phi^{(0)}(x)$ plays a role of the sources that generate the correlators with CFT operators $\mathcal{O}(x)$. 
To define the wave functional of universe $\Psi_{\text{dS}}$, we have to also impose some initial condition which will be mentioned below. 

A distinct point from the AdS/CFT is that the boundary $\mathcal{I}^+$ where CFT lives does not include the bulk time, since it is a space-like surface.
Therefore even if the gravity theory is unitary, the dual CFT is not necessarily restricted to be unitary theory. 
Indeed, the analytic continuation from AdS to dS: $L_{\text{AdS}}\to iL_{\text{dS}}$ implies the central charge of the dual CFT
\begin{align}
    c \, \sim \, i^{d-1}\frac{L_{\text{dS}}^{d-1}}{G_N^{(d+1)}}
\end{align}
does not generally satisfy the unitarity condition. For example, in the $d=3$ case the central charge takes negative value. In this case, a concrete example is known \cite{Anninos:2011ui}, where the four-dimensional higher-spin dS gravity is dual to the large $N$ limit of the $Sp(N)$ model, and the latter theory is known to be non-unitary.
This duality is obtained by the analytic continuation from the Klebanov-Polyakov duality \cite{Klebanov:2002ja}. 

In \cite{Hikida:2021ese,Hikida:2022ltr}, $d=2$ case has been investigated and CFT$_2$ dual to Einstein gravity with positive cosmological constant is proposed to be non-chiral $SU(2)_{\hat{k}}$ Wess-Zumino-Witten (WZW) model, while the semi-classical limit of the gravity side corresponds to a somewhat unfamiliar limit of the level $\hat{k}$ as we'll describe shortly. 
Since the central charge of CFT$_2$ is
\begin{align}
    c=i\frac{3L_{\text{dS}}}{2G_N}\equiv ic^{(g)},
\end{align}
the limit of $\hat{k}$ that corresponds to the large $c^{(g)}$ limit through the well-known formula $c=3\hat{k}/(\hat{k}+2)$ can be found as 
\begin{align}
    \hat{k}\to-2+\frac{6i}{c^{(g)}}+\mathcal{O}\left(\frac{1}{c^{(g)2}}\right).
\end{align}
Therefore the semi-classical limit in the gravity side corresponds to the critical level limit $\hat{k}\to-2$.
The extension to higher-spin gravity is easily accomplished by replacing $SU(2)_{\hat{k}}$ with $SU(N)_{\hat{k}}$, then the semi-classical limit becomes to $\hat{k}\to -N$. 
This duality has an ambiguity of including additional matter fields, though these does not contribute to quantities in the leading order (i.e. tree-level). 
An example with including matter fields is given by the analytic continuation of the Gaberdiel-Gopakumar duality \cite{Gaberdiel:2010pz,Gaberdiel:2012ku} (see also \cite{Gaberdiel:2012uj,Perlmutter:2012ds}) that relates three-dimensional AdS higher-spin gravity including a specific scalar field to the coset model 
\begin{align}
    \frac{SU(N)_{\hat{k}}\times SU(N)_1}{SU(N)_{\hat{k}+1}}.
\end{align}
A special case with $N=2$ gives the Einstein gravity. Note that in \cite{Hikida:2021ese,Hikida:2022ltr}, the initial state for $\Psi_{\text{dS}}$ is assumed to satisfy the Hartle-Hawking's no-boundary condition \cite{Hartle:1983ai}, whose semi-classical limit can be realized as the Euclidean path integral on the hemisphere $B^3$. Recently some justifications for taking the Hartle-Hawking state were discussed in \cite{Witten:2021nzp,Chen:2023prz,Chen:2023sry}.

Although the conjecture was given for Lorentzian dS$_3$, some calculations in \cite{Hikida:2021ese,Hikida:2022ltr} are done for Euclidean dS$_3$, i.e. a three-sphere $S^3$. 
A derivation of the Euclidean version is as follows (see also the introduction of \cite{Hikida:2022ltr}). The partition function of $S^3$ is constructed as the squared norm of the wave functional of universe:
\begin{align}\label{S3pf}
    Z_{\text{G}}[S^3]=\int \mathcal{D}g^{(0)}_{\mu\nu}\left|\Psi_{\text{dS}}\left[g^{(0)}_{\mu\nu}\right]\right|^2.
\end{align}
In particular, $\Psi_{\text{dS}}$ approximates in the semi-classical regime as 
\begin{align}
    \Psi_{\text{dS}}[S^2]\sim \exp\left(-I^{(\text{E})}_{\text{G}}[B^3]+i I^{(\text{L})}_{\text{G}}[\text{dS}_{3}^{T>0}]\right).
\end{align}
Here $I^{(\text{E})}_{\text{G}}[B^3]$ is the contribution from the on-shell action for the Euclidean part of $\Psi_{\text{dS}}$ at $T=0$, interpreted as the Gibbons-Hawking de Sitter entropy \cite{Gibbons:1976ue,Gibbons:1977mu}. The imaginary part $I^{(\text{L})}_{\text{G}}[\text{dS}_{3}^{T>0}]$ is the contribution from the on-shell action for the Lorentzian part, i.e., the future half $T>0$ of $\text{dS}_3$. In this approximation, the partition function on $S^3$ \eqref{S3pf} is expressed as\footnote{Note that the notation of $Z_{\text{CFT}}$ in this paper is different from that in \cite{Hikida:2022ltr} because the proposed dual CFT in this paper is a double copy of $SU(2)$ non-chiral WZW models.}
\begin{align}
    Z_{\text{G}}[S^3]\sim \left|\Psi_{\text{G}}[S^2]\right|^2=\left|Z_{\text{WZW}}^{SU(2)}\right|^2 \, \equiv \, Z_{\textrm{CFT}}.
\end{align}

Therefore we can expect that the Euclidean dS$_3$ gravity theory on $S^3$ is dual to a double copy of the CFT$_2$ that itself is dual to Lorentzian dS$_3$. This relation would imply that the holographic duality can be applied to a hemisphere; splitting the three-sphere as $S^3\simeq B^3\cup \overline{B^3}$, the Euclidean dS$_3$ gravity on each $B^3$ is equivalent to a CFT on $\partial B^3= S^2$. A direct construction of this statement is attempted in Appendix \ref{CSonB3}.


\section{Second order formalism}
\label{sec:3d gravity}
In this section, we study the three-dimensional gravity discussed in \cite{Hikida:2021ese, Hikida:2022ltr} in the second order formalism,
and we will show that the its classical contribution can be completely captured by a JT gravity on $S^2$ with a non-standard boundary term.

The action of the 3d gravity on $S^3$ we study is given by
	\begin{align}
		I_G \, = \, - \frac{1}{16 \pi G_N} \int d^3x \sqrt{g_3} \, \big( R_3 - 2 \big) \, .
	\label{I_G}
	\end{align}
In the rest of this paper, we fix the dS radius as $L_{\text{dS}}=1$. The black hole solution is given by
	\begin{align}
		ds^2 \, = \, (r_0^2 - r^2) d\t^2 \, + \, \frac{dr^2}{r_0^2 - r^2} \, + \, r^2 d\vp^2 \, ,
	\label{metric}
	\end{align}
where $r_0 = \sqrt{1-8 G_N E_j}$ in terms of the notation of \cite{Hikida:2021ese, Hikida:2022ltr}.
The vacuum solution corresponds to $E_j=0$ (i.e. $r_0 = 1$), while in the following we use general $r_0 \ge 1$.
The periodicity of the Euclidean time is 
	\begin{align}
		\t \, \sim \, \t \, + \b \, ,
	\end{align}
and the black hole inverse temperature is given by
	\begin{align}
		\b \, = \, \frac{2\pi}{r_0} \, .
	\end{align}
From this expression, we can see that the horizon location $r_0$ is proportional to the black hole temperature $\b^{-1}$.
The classical on-shell action was computed in \cite{Hikida:2021ese, Hikida:2022ltr} as
	\begin{align}
		I_G^{\onshell} \, = \, - \frac{r_0^2 \, \b}{4G_N} \, = \, - \frac{\pi \, r_0}{2G_N} \, .
	\label{I_onshell}
	\end{align}

If we consider the near-horizon geometry $r \to r_0$, then the metric (\ref{metric}) becomes
	\begin{align}
		ds^2 \, \approx \, (r_0^2 - r^2) d\t^2 \, + \, \frac{dr^2}{r_0^2 - r^2} \, + \, r_0^2 d\vp^2 \, ,
	\end{align}
so the near-horizon geometry is $S^2 \times S^1$. 
Furthermore, if we imagine that we could take a low temperature limit $r_0 \to 0$ (even though originally the horizon location is restricted in $r_0 \ge 1$),
the 3d physics can be approximated by taking the zero mode contribution in the $S^1$ direction.
In particular, the on-shell action (\ref{I_onshell}) is just linear in $r_0$ (i.e. black hole temperature), we expect the on-shell action can be recovered from the above approximation.
We will show that this is indeed the case in the following subsection.

Before proceeding to the dimensional reduction, we also review the two-linked Wilson line geometry.
The background metric for this geometry is
	\begin{align}
		ds^2 \, = \, (r_0^2 - r^2) d\wtd{\t}^2 \, + \, \frac{dr^2}{r_0^2 - r^2} \, + \, r^2 d\vp^2 \, ,
	\end{align}
with
	\begin{align}
		\wtd{\t} \, \sim \, \wtd{\t} \, + \wtd{\b} \, , \qquad \quad \wtd{\b} \, = \, \frac{2\pi r_*}{r_0} \, , 
	\end{align}
where $r_* = \sqrt{1-8 G_N E_l}$ in terms of the notation of \cite{Hikida:2021ese, Hikida:2022ltr}.
Then the classical on-shell action for this geometry is given by
	\begin{align}
		I_G^{\onshell} \, = \, - \frac{r_0^2 \, \wtd{\b}}{4G_N} \, = \, - \frac{\pi r_0 r_*}{2G_N} \, .
	\label{I_onshell_alpha}
	\end{align}
In this case, we also have the on-shell action proportional to the black hole temperature, so the dimensional reduction is possible.

\subsection{Dimensional Reduction to 2D}
\label{sec:jt}
Now we study the dimensional reduction from the 3D gravity described above.
We consider the following dimensional reduction ansatz for the 3D metric: 
	\begin{align}
		ds_3^2 \, = \, g^{(2)}_{\m\n} \, dx^\m dx^\n \, + \, \Phi^2 d\vp^2 \, ,
	\label{ansatz}
	\end{align}
where $g^{(2)}_{\m\n}$ and $\Phi$ are functions of $x^\m$ and independent of $\vp$.
With this ansatz, the 3D Ricci scalar is reduced as
	\begin{align}
		R_3 \, = \, R_2 \, - \, 2 \P^{-1} \Box_2 \P \, ,
	\end{align}
where $\Box_2$ is the Laplacian corresponding to $g^{(2)}_{\mu\nu}$.
Using this result, the 3D action is reduced to 
	\begin{align}
		I_G \, = \, - \frac{1}{8 G_N} \int d^2x \sqrt{g_2} \, \P \big( R_2 - 2 \big) \, + \, \frac{1}{4G_N} \int_0^\b d\t \Big[ \sqrt{g_2} \, g_2^{rr} \, \pa_r \P \Big]^{r=r_0}_{r=0} \, .
	\label{I_JT}
	\end{align}
The 2D bulk action is the JT action, but we note that the boundary term is not the ordinary Dirichlet boundary term appears in AdS$_2$ \cite{Maldacena:2016upp} or Lorentzian dS$_2$ \cite{Maldacena:2019cbz}.
In fact, all possible boundary conditions are classified in \cite{Goel:2020yxl}, and according to their terminology, our boundary term corresponds to the $NN^*$ boundary condition.
This roughtly corresponds to imposing Neumann boundary condition for the metric and Neumann boundary condition for the dilaton.
In appendix~\ref{app:bc_jt}, we will explain details of the variational principle for this JT action and its boundary conditions.

The equations of motion of this 2d action are
	\begin{gather}
		R_2 \, = \, 2 \, , \\
		\nabla_\m \nabla_\n \P \, - \, g_{\m\n} \nabla^2 \P \, - \, \P \, g_{\m\n} \, = \, 0 \, .
    \label{dilaton-eq}
	\end{gather}
The corresponding static background solution is given by
	\begin{align}
		ds_2^2 \, &= \, (r_0^2 - r^2) d\t^2 \, + \, \frac{dr^2}{r_0^2 - r^2} \, , \\
		\P \, &= \, r \, .
	\label{2D_metric}
	\end{align}
This solution agrees with the original 3D background solution (\ref{metric}) provided with the ansatz (\ref{ansatz}).
For the dilaton solution, we fixed the coefficient in order to match with the 3D solution (\ref{metric}).
Namely, we imposed the boundary condition
	\begin{align}
		\pa_r \P(r = r_0) \, = \, 1 \, .
	\end{align}
With this background solution, now we would like to compute the classical on-shell action.
Since the background satisfies $R_2 \, = \, 2$, the 2D bulk action does not give any contribution.
The on-shell action is given by the boundary term as
	\begin{align}
		I_G^{\onshell} \, &= \, \frac{\b}{4 G_N} \Big[ \sqrt{g_2} \, g_2^{rr} \, \pa_r \P \Big]^{r=r_0}_{r=0} \nn\\[2pt]
		&= \, \frac{\b}{4 G_N} \Big[ r_0^2 - r^2 \Big]^{r=r_0}_{r=0} \nn\\[2pt]
		&= \, - \frac{\pi \, r_0}{2G_N} \, .
	\label{I_onshell_2D}
	\end{align}
This agrees with (\ref{I_onshell}).
The above computation shows that even though we are dealing with $S^2$ geometry,
in order to recover the on-shell action obtained in the 3D gravity, we need to include some boundary terms at $r=0$ and $r=r_0$.
This situation is analogous to the computation of the on-shell action from the Chern-Simons theory discussed in section~3.1 of \cite{Hikida:2022ltr}.
Similar ideas were also discussed for Lorenzian dS case in \cite{Susskind:2021omt, Susskind:2021dfc, Kawamoto:2023nki}.

For the two-linked Wilson line geometry, with the ansatz (\ref{ansatz}), the reduction of the 3D Ricci scalar contains an additional delta function which is responsible for the defect at $r=0$:
	\begin{align}
		R_3 \, = \, R_2 \, - \, 2 \P^{-1} \Box_2 \P \, - \, 2\a \d^2(x) \, ,
	\end{align}
where $\a=2\pi(1-r_*)$. Therefore, the resulting 2D action is the defect JT gravity:
	\begin{align}
		I_G \, = \, - \frac{1}{8 G_N} \left[ \int d^2x \sqrt{g_2} \, \P \big( R_2 - 2 \big) -2\a \P(0) \right] \, + \, \frac{1}{4G_N} \int_0^{\wtd{\b}} d\wtd{\t} \Big[ \sqrt{g_2} \, g_2^{rr} \, \pa_r \P \Big]^{r=r_0}_{r=0} \, .
	\end{align}
We note that there is an extra term $2\a \P(0)$ in the bulk action inserted at $r=0$. Also now the Eulidean time is given by $\wtd{\t}$ with period $\wtd{\b}$.
This theory was studied in \cite{Mertens:2019tcm, Maxfield:2020ale, Witten:2020wvy, Mefford:2020vde},
and the background solution is given by the same form as in (\ref{2D_metric}) except the $\t$ is replaced by $\wtd{\t}$.
Therefore, the evaluation of the on-shell action is also parallel to (\ref{I_onshell_2D}) except the $\b$ coming from the $\t$ integral is now replaced by $\wtd{\b}$ as
	\begin{align}
		I_G^{\onshell} \, = \, - \frac{r_0^2 \, \wtd{\b}}{4G_N} \, = \, - \frac{\pi r_0 r_*}{2G_N} \, .
	\end{align}
This agrees with (\ref{I_onshell_alpha}).

Hence in this section, we have shown that the classical dynamics of the 3D gravity is completely captured by the JT gravity in 2D with the $NN^*$ boundary condition.

\section{First order formalism}
\subsection{Chern-Simons gravity}
\label{sec:chern-simons}
We can also discuss the dimensional reduction in the first order formalism.
However, before discussing the dimensional reduction in section~\ref{sec:bf}, in this subsection we are going to clarify some subtleties about bounder terms in the Chern-Simons gravity.
In the following, we denote the three-dimensional indices by capital letters and two-dimensional indices by lowercase.
$M, N, \m, \n$ are spacetime indices contracted by $g_{MN}$ or $g_{\m\n}$ and $A, B, a, b$ are tangent space indices contracted by $\d_{AB}$ or $\d_{ab}$.
In terms of the frame field $E_M{}^A$ defined by $g_{MN} = E_M{}^A E_N{}^B \d_{AB}$ and the spin connection $\O_M{}^A$, the Einstein-Hilbert action
	\begin{align}
		I_G \, = \, - \frac{1}{16 \pi G_N} \int d^3x \sqrt{g_3} \, \big( R_3 - 2 \L \big) \, .
	\end{align}
is written as
	\begin{align}
		I_G \, = \, - \frac{1}{16 \pi G_N} \int d^3x \, \e^{MNK} \left( E_{MA} R^A{}_{NK} \, - \, \frac{\L}{3} \e_{ABC} E_M{}^A E_N{}^B E_K{}^C \right) \, ,
	\end{align}
where we recovered the cosmological constant $\L$ to include the AdS case as well. (We will review the double Wick rotation from AdS$_3$ to $S^3$ in appendix~\ref{app:rotation}.)
Now introducing the gauge fields by
	\begin{align}
		A_M^{\pm A} \, := \, - \big( \O_M{}^A \, \pm \, \sqrt{\L} \, E_M{}^A \big) \, .
	\label{gaugeA}
     \end{align}
the Einstein-Hilbert action is written in terms of the Chern-Simons actions as \cite{Achucarro:1986uwr, Witten:1988hc}
	\begin{align}
		I_G \, = \, I_{\textrm{CS}}[A^+] \, - \, I_{\textrm{CS}}[A^-] \, + \, I_{\textrm{bdy}}[A^+, A^-] \, ,
	\label{non-chiral action}
	\end{align}
where
	\begin{align}
		&I_{\textrm{CS}}[A] \, = \, - \frac{k}{8\pi} \int d^3x \, \e^{MNK} \left[ A_M^A \pa_N A_{KA} \, + \, \frac{\e_{ABC}}{3} \, A_M^A A_N^B A_K^C \right] \, , \label{CS_action} \\
		&I_{\textrm{bdy}}[A^+, A^-] \, = \, - \frac{k}{8\pi} \int d^2x \, \big( A^{+A}_\t A^-_{\vp A} - A^{+A}_\vp A^-_{\t A} \big) \, , \label{bdy_action1}
    \end{align}
and 
	\begin{align}
		k \, = \, \frac{1}{4G_N\sqrt{\L}} \, .
	\label{level-k}
    \end{align}

For the case of $S^3$ ($\L=1$), since $SO(4) \simeq SU(2) \times SU(2)$, we can introduce the $SU(2)$ generators $J_A$.
(For explicit representations and their properties, see appendix~\ref{app:su(2)}.)
Then, if we define
	\begin{align}
		A_M \, = \, A_M^A \, J_A \, ,
	\end{align}
the Chern-Simons action becomes
	\begin{align}
		I_{\textrm{CS}}[A] \, = \, \frac{k}{4\pi} \int d^3x \, \e^{MNK} \tr \left[ A_M \pa_N A_K \, + \, \frac{2}{3} \, A_M A_N A_K \right] \, .
	\label{I_CS}
    \end{align}

The background solutions which give the metric (\ref{metric}) are
	\begin{align}
		E^A \, &= \, \big\{ \sqrt{r_0^2 - r^2} \, d\t \, , \, \frac{dr}{\sqrt{r_0^2 - r^2}} \, , \, r d\vp \big\} \, , \\
		\O^A \, &= \, \big\{ - \sqrt{r_0^2 - r^2} \, d\vp \, , \, 0 \, , \, -r d\t \big\} \, .
	\end{align}
Therefore, on the basis of $SU(2)$ (\ref{SU(2)_generators}), the background solutions of the gauge fields are
	\begin{align}
		A^{\pm} \, = \, \begin{pmatrix} \frac{\mp i dr}{2\sqrt{r_0^2 - r^2}} & \frac{i r \pm \sqrt{r_0^2 - r^2}}{2} \, dx^{\mp} \\[8pt]
		\frac{i r \mp \sqrt{r_0^2 - r^2}}{2} \, dx^{\mp} & \frac{\pm i dr}{2\sqrt{r_0^2 - r^2}} \end{pmatrix} \, ,
	\label{gauge_solution}
    \end{align}
where $x^{\pm} = \t \pm \vp$.
With this background solutions, we can evaluate the on-shell action as
	\begin{align}
		I_{\textrm{CS}}^{\onshell}[A^\pm] \, &= \, 0 \, , \\
		I_{\textrm{bdy}}^{\onshell}[A^+, A^-] \, &= \, - \frac{\b r_0^2}{4G_N} \, = \, - \frac{\pi r_0}{2G_N} \, .
	\label{I_on-shell}
    \end{align}
This agrees with (\ref{I_onshell}).

On the other hand, the Chern-Simons action (\ref{I_CS}) can be rewritten as
	\begin{align}
		I_{\textrm{CS}}[A] \, &= \, \frac{k}{4\pi} \int d^3x \, \tr\Big[ A_\vp \dot{A}_r - A_r \dot{A}_\vp + 2A_\t F_{r\vp} \Big] \nn\\
		&\qquad\, - \, \frac{k}{4\pi} \int_{r=r_0} d^2x \, \tr \big[ A_\t A_\vp \big] \, + \, \frac{k}{4\pi} \int_{r=0} d^2x \, \tr \big[ A_\t A_\vp \big] \, ,
    \label{I_CS2}
	\end{align}
where the dot denotes a derivative with respect to $\t$. Therefore, the variation of the Chern-Simons action with respect to $A_\t$ is
	\begin{align}
		\d I_{\textrm{CS}}[A] \, = \, (\textrm{EOM}) \, - \, \frac{k}{4\pi} \int_{r=r_0} d^2x \, \tr \big[ \d A_\t A_\vp \big] \, + \, \frac{k}{4\pi} \int_{r=0} d^2x \, \tr \big[ \d A_\t A_\vp \big] \, ,
	\end{align}
while the variation of the boundary term (\ref{bdy_action1}) is 
	\begin{align}
		\d I_{\textrm{bdy}}[A^+, A^-] \, = \, \frac{k}{4\pi} \int_{r=r_0} d^2x \, \tr \Big[ \d A^+_\t A^-_\vp - \d A^-_\t A^+_\vp \Big] \, - \, \frac{k}{4\pi} \int_{r=0} d^2x \, \tr \Big[ \d A^+_\t A^-_\vp - \d A^-_\t A^+_\vp \Big] \, .
	\end{align}
Therefore, the boundary action (\ref{bdy_action1}) imposes boundary conditions
	\begin{align}
		A^+_\vp \, = \, A^-_\vp \, ,
	\end{align}
at $r=r_0$ and $r=0$.
This is obviously not consistent with (\ref{gauge_solution}), so we need to replace the boundary action (\ref{bdy_action1}) by another one.
This is the same situation as in the Chern-Simons gravity in AdS$_3$ \cite{Coussaert:1995zp, Donnay:2016iyk}, and our prescription is also the same as in the AdS$_3$ case.
In AdS$_3$, one eliminates the second line of (\ref{I_CS2}) and include an additional boundary action proportional to $(A_\vp^+)^2+(A_\vp^-)^2$.
Since we are dealing with Euclidean case, $A_\vp$ is equivalent to $A_\t$ here, so we employ a boundary action proportional to $(A_\t^+)^2+(A_\t^-)^2$.
We also keep the Chern-Simons action as written in (\ref{I_CS2}), so we add extra terms in our boundary action in order to subtract the second line of (\ref{I_CS2}).
Hence our new boundary action is 
	\begin{align}
		\wtd{I}_{\textrm{bdy}}[A^+, A^-] \, = \, \wtd{I}_{\textrm{bdy}}^{(r=r_0)}[A^+, A^-] \, + \, \wtd{I}_{\textrm{bdy}}^{(r=0)}[A^+, A^-] \, ,
	\end{align}
with
	\begin{align}\label{bdy1}
		\wtd{I}_{\textrm{bdy}}^{(r=r_0)}[A^+, A^-] \, &= \, \frac{k}{4\pi} \int_{r=r_0} d^2x \, \tr \Big[ - A_\t^+ A_\vp^+ - A_\t^+ A_\t^+ + A_\t^- A_\vp^- - A_\t^- A_\t^- \Big] \, , \\
	\label{bdy2}	\wtd{I}_{\textrm{bdy}}^{(r=0)}[A^+, A^-] \, &= \, - \frac{k}{4\pi} \int_{r=0} d^2x \, \tr \Big[  A_\t^+ A_\vp^+ - A_\t^- A_\vp^- \Big] \, .
	\end{align}
This new boundary action imposes no boundary condition at $r=0$ and imposes $A^+_+ = 0 = A^-_-$ at $r=r_0$.
With the background solution (\ref{gauge_solution}), the on-shell value of this new boundary term gives
	\begin{align}
		\wtd{I}_{\textrm{bdy}}^{\onshell}[A^+, A^-] \, = \, - \frac{\b r_0^2}{8G_N} \, = \, - \frac{\pi r_0}{4G_N} \, .
	\end{align}
This is actually half of the on-shell action we had in section~\ref{sec:3d gravity}.
This means that by changing the boundary action, we have changed our theory, so that we now need two-copies of the Chern-Simons action we have been discussing here.


\subsection{BF theory}
\label{sec:bf}
In this subsection, we discuss the dimensional reduction in the first order formalism.
A dimensional reduction of a chiral Chern-Simons action was previously discussed in \cite{Achucarro:1992mb},
while in this section, we study a dimensional reduction of the non-chiral Chern-Simons action, which leads to the BF action.
BF theory with $SL(2; \mathbb{R})$ gauge symmetry was previously discussed for example in \cite{Isler:1989hq, Chamseddine:1989yz, Iliesiu:2019xuh}.

\subsubsection{Dimensional reduction}
In first order formalizm, the ansatz \eqref{ansatz} is expressed by a vielbein $E$ with components
\begin{align}\begin{aligned}
    E^{A} \, &= \, e^{a} \qquad (A=a=1,2), \\
    E^{3} \, &= \, \Phi(x)\, d\varphi,
\end{aligned}\end{align}
where $e^a=e^a{}_\m dx^\m$ is a 2d vielbein associated with 2d metric $g^{(2)}_{\mu\nu}$. 
The spin connection $\Omega^A$ is now given by
\begin{align}
    \begin{aligned}
        \Omega^1&=-e_2^{~\mu}\partial_\mu\Phi\,d\varphi,\\
        \Omega^2&=e_1^{~\mu}\partial_\mu\Phi\,d\varphi,\\
        \Omega^3&=\omega,
    \end{aligned}
\end{align}
where $\omega$ is the component $\omega=\omega^1_{~2}$ of a spin connection in two dimensions. 
These forms indeed satisfy the torsion-free conditions $dE^A-\epsilon_{ABC}\Omega^B\wedge E^C=0$. Then the Chern-Simons gauge connections $A^{\pm A}$ are given by
\begin{align}\begin{aligned}
    A^{+1}&=e_2^{~\mu}\partial_\mu\Phi\,d\varphi-e^1\qquad& A^{-1}&=e_2^{~\mu}\partial_\mu\Phi\,d\varphi+e^1\\
    A^{+2}&=-e_1^{~\mu}\partial_\mu\Phi\,d\varphi-e^2& A^{-2}&=-e_1^{~\mu}\partial_\mu\Phi\,d\varphi+e^2\\
    A^{+3}&=-\omega-\Phi\,d\varphi,&A^{-3}&=-\omega+\Phi\,d\varphi
\end{aligned}\end{align}

Let us evaluate the Chern-Simons gravity action 
\begin{align}
    I_{\text{G}}=I_{\text{CS}}[A^+]-I_{\text{CS}}[A^-]
\end{align}
without any boundary terms.
The Lagrangian density is calculated as 
\begin{align*}
    &\frac{k}{4\pi}\tr\left[A^+\wedge dA^++\frac{2}{3}A^+\wedge A^+\wedge A^+\right]-\frac{k}{4\pi }\tr\left[A^-\wedge dA^-+\frac{2}{3}A^-\wedge A^-\wedge A^-\right]\\
    &=\frac{1}{16\pi G_N}\left[\Box_2\Phi e^1\wedge e^2\wedge d\varphi -2\Phi d\omega\wedge d\varphi+\frac{2}{L^2}\Phi e^1\wedge e^2\wedge d\varphi\right]
\end{align*}
By performing the $\varphi$-integral and utilizing the properties 
\begin{align}
    e^1\wedge e^2=\sqrt{g_2}d^2x, \quad d\omega=\frac{1}{2}R_2\sqrt{g_2}d^2x,
\end{align}
the action has the form 
\begin{align}\begin{aligned}\label{CSredbulk}
    I_{\text{bulk}}&=\frac{1}{8 G_N}\int_{M_2} \Box_2\Phi e^1\wedge e^2-2\Phi d\omega+2\Phi e^1\wedge e^2\\
    &=-\frac{1}{8G_N}\int_{M_2}d^2x\sqrt{g_2}\Phi(R_2-2)+\frac{1}{8G_N}\int_{M_2}d^2x\sqrt{g_2}\Box_2 \Phi
\end{aligned}\end{align}
Next we consider the original boundary term given by \eqref{bdy_action1}.
This boundary action is reduced to 
\begin{align}\begin{aligned}\label{CSredbulk}
    I_{\text{bdy}} \, = \, \frac{1}{8G_N}\int_{M_2}d^2x \, \sqrt{g_2}\Box_2 \Phi \, ,
\end{aligned}\end{align}
so that the total action agrees with (\ref{I_JT}).
Finally we consider the modified boundary term given by \eqref{bdy1} and \eqref{bdy2}. Assuming $g^{(2)}$ is the diagonal metric, the boundary term at $r=0$ takes the form 
\begin{align}
    \tilde{I}_{\text{bdy}}^{(r=0)}
    =-\frac{1}{8G_N}\int_{r=0}d\tau(e^2_\tau\partial_1\Phi-e^1_\tau\partial_2\Phi +\omega_\tau\Phi)
    =-\frac{1}{8G_N}\int d^2x\,\Box_2\Phi.
\end{align}
Since these terms cancel with the additional terms in \eqref{CSredbulk}, this boundary term imposes no boundary condition at $r=0$. The boundary term at $r=r_0$ terns out to be 
\begin{align}\label{bdyr0}
    \tilde{I}_{\text{bdy}}^{(r=r_0)}=-\frac{1}{8G_N}\int_{r=r_0}d\tau(e^2_\tau\partial_1\Phi-e^1_\tau\partial_2\Phi+\omega_\tau\Phi)-\frac{1}{8G_N}\int_{r=r_0}d\tau\left(g^{(2)}_{\tau\tau}+\omega_\tau^2\right).
\end{align}
The sum of the bulk and boundary terms, we have 
\begin{align}
    I_{\text{G}}=-\frac{1}{8G_N}\int_{M_2}d^2x\sqrt{g_2}\Phi\left(R_2-2\right)-\frac{1}{8G_N}\int_{r=r_0}d\tau(g^{(2)}_{\tau\tau}+\omega_\tau^2).
\end{align}

We note that the boundary term differs from the one discussed in section~\ref{sec:jt}.
This is not surprise since we changed the boundary term in the Chern-Simons gravity in section~\ref{sec:chern-simons}.
This means that we have changed the theory. 
In the current case, we have $g^{(2)}_{\tau\tau}=r_0^2 - r^2$ and $\omega_\tau^2=-r$, so the classical on-shell action is now given by
	\begin{align}
		I_G^{\onshell} \, = \, - \frac{r_0^2 \, \b}{8G_N} \, = \, - \frac{\pi \, r_0}{4G_N} \, .
	\end{align}
This on-shell action is half of that of the original 3d gravity theory.
Hence as we discussed in section~\ref{sec:chern-simons}, with this boundary term, we need twice of the current theory in order to match with the original theory.




Next let us split the gauge connection $A^+$ as 
\begin{align}
    A^+=a^++b^+\, d\varphi,
\end{align}
where $a$ and $b$ are a 1-form and 0-form, respectively, given by 
\begin{align}\begin{aligned}\label{ab}
    a^+&=e^1J_1+e^2J_2-\omega J_3,\\
    b^+&=e_2^{~\mu}\partial_\mu\Phi J_1-e_1^{~\mu}\partial_\mu\Phi J_2+\Phi J_3.
\end{aligned}\end{align}
We can easily deduce that the Chern-Simons action reduces to the BF theory:
\begin{align}
    I_{\text{CS}}[A^+]&=\frac{k}{4\pi}\int_{M_3} \tr\left[2b^+da^++2b^+a^+\wedge a^+-d(b^+a^+)\right]\wedge d\varphi \\
    &=k\int_{M_2}\tr[b^+f^+]-\frac{k}{2}\int_{M_2}d\tr[b^+a^+],
\end{align}
where $f^+\equiv da^++a^+\wedge a^+$. 
Next we consider the second part $I_{\text{CS}}[A^-]$. The gauge field $A^-$ is split as 
\begin{align}
    A^-=a^-+b^-d\phi,
\end{align}
where 
\begin{align}\begin{aligned}\label{abbar}
    a^-&=-e^1J_1-e^2J_2-\omega J_3,\\
    b^-&=e_2^{~\mu}\partial_\mu\Phi J_1-e_1^{~\mu}\partial_\mu\Phi J_2-\Phi J_3.
\end{aligned}\end{align}
The dimensional reduction of Chern-Simons action can be done in the same way above:
\begin{align}
    I_{\text{CS}}[A^-]=k\int_{M_2}\tr[b^-f^-]-\frac{k}{2}\int_{M_2}d\tr[b^-a^-],
\end{align}
where $f^-\equiv da^-+a^-\wedge a^-$. Comparing the expressions \eqref{ab} and \eqref{abbar}, we can find that $\tr[b^-f^-]=-\tr[b^+f^+],\ \tr[b^-a^-]=-\tr[b^+a^+]$. Therefore 
\begin{align}\label{BFnonchiral}
    I_{\text{CS}}[A^+]-I_{\text{CS}}[A^-]=2k\int_{M_2}\tr[b^+f^+]-k\int_{M_2}d\tr[b^+a^+].
\end{align}

The boundary term that reproduces \eqref{bdyr0} is given by
\begin{align}
    I_{\text{bdy}}=k\int_{r=r_0}\tr\left[a^+b^++(a^+_\tau)^2 d\tau\right].
\end{align}
Therefore in terms of BF theory the total action can be found to be 
\begin{align}
    I_{\text{G}}=2k\int \tr[b^+f^+]+k\int_{r=r_0}d\tau\tr\left[(a^+_\tau)^2\right].
\end{align}


\subsubsection{BF theory analysis}
In this subsection, let us start with the action obtained in the previous subsection
\begin{align}\label{su2bf}
    I_{\text{G}}=2k\int_{M_2}\tr[bf]+\int_{r=r_0}d\tau\tr[a^2_\tau]
\end{align}
and solve the theory. Henceforth we omit the superscript $\pm$ since $I_{\text{G}}$ has the form of a single BF theory.
We parameterize the components of $a$ and $b$ as 
\begin{align}
    a&=e^{1} J_1+e^2 J_2+ \omega J_3 \\
    b&=\Phi_1 J_1+\Phi_2J_2+\Phi J_3
\end{align}
The bulk EOM on $b$ imposes the flat connection 
\begin{align}
    f=0.
\end{align}
The bulk EOM $db+[a,b]=0$ with respect to $a$ is written in terms of components as
\begin{align}\begin{aligned}
    \partial_\mu \Phi&=-e^1_\mu \Phi_2+e^2_\mu\Phi_1,\\
    \partial_\mu\Phi_1&=\omega_\mu\Phi_2-e^2_\mu\Phi,\\
    \partial_\mu\Phi_2&=-\omega_\mu\Phi_1+e^1_\mu\Phi,
\end{aligned}\end{align}
From the above equations we can find $\Phi$ satisfies 
\begin{align}
    \nabla_\mu\nabla_\nu\Phi=-g_{\mu\nu}\Phi,
\end{align}
where $\nabla_\mu$ is the covariant derivative, and
\begin{align}\begin{aligned}
    \Phi_1&=e_2^{\mu}\partial_\mu\Phi,\\
    \Phi_2&=-e_1^{\mu}\partial_\mu\Phi
\end{aligned}\end{align}

Let us consider the background solution
\begin{align}
    e^1=\sqrt{r_0^2-r^2}d\tau,\quad e^2=\frac{dr}{\sqrt{r_0^2-r^2}},\quad \Phi=r,
\end{align}
then the remaining fields are determined by 
\begin{align}
    \omega=-rd\tau,\quad \Phi_1=\sqrt{r_0^2-r^2},\quad \Phi_2=0.
\end{align}
Therefore $a$ and $b$ become
\begin{align}
    a&=\frac{1}{2}
    \begin{pmatrix}
        \frac{idr}{\sqrt{r_0^2-r^2}} & -(\sqrt{r_0^2-r^2}+ir)d\tau \\
        (\sqrt{r_0^2-r^2}-ir)d\tau & \frac{-idr}{\sqrt{r_0^2-r^2}},
    \end{pmatrix}\\
    b&=\frac{1}{2}
    \begin{pmatrix}
        0 & -\sqrt{r_0^2-r^2}+ir\\
        \sqrt{r_0^2-r^2}+ir & 0
    \end{pmatrix}.
\end{align}

Since the bulk term vanishes on shell due to $f=0$, the on-shell action only comes from the boundary term. On the above solution, the boundary term takes
\begin{align}
    I_{\text{G}}^{\text{on-shell}}=I_{\text{bdy}}^{\text{on-shell}}=-\frac{\pi  r_0}{4G_{N}},
\end{align}
which is consistent with the 3d calculations of the on-shell action.

\subsubsection{Higher-spin extension}
One advantage of adopting the first order formalism is that the generalization to higher-spin gravity is straightforward, just replacing the gauge group $SU(2)$ with $SU(N)$ \cite{Vasiliev:1995sv,Alekseev:1988ce}. From the discussions of Chern-Simons formulation of three-dimensional higher-spin gravity theory, the Chern-Simons level is related to the gravitational notions as
\begin{align}
    k=\frac{1}{8G_N\epsilon_N},
\end{align}
where $\epsilon_N\equiv N(N^2-1)/12$.
First we describe the classification of the smooth solutions to the bulk equations of motion
\begin{align}\label{BFEOM}
    f=0,\quad db+[a,b]=0.
\end{align}
Here we consider the principal embedding of $SU(2)$ into $SU(N)$, where the $SU(2)$ generators $J_a$ embedded satisfy
\begin{align}
    [J_a,J_b]&=\epsilon_{abc}J_c, \\
    \tr[J_aJ_b]&=\epsilon_N\delta_{a,b},\qquad \epsilon_N=\frac{1}{12}N(N^2-1).
\end{align}
We classify the smooth flat connections in the same way as \cite{Castro:2011iw}, where the trivial holonomy is imposed as the smoothness condition. Expressing the connection as 
\begin{align}
    a=h^{-1}a_\tau h\,d\tau+h^{-1}dh,\qquad h=e^{rJ_1},
\end{align}
the holonomy takes the form 
\begin{align}
    \text{Hol}_\tau(a)=h^{-1}\mathcal{P}\exp\left(\int_0^\beta d\tau a_\tau\right)h.
\end{align}
Imposing the trivial holonomy condition, which is given by $\text{Hol}(A)=\mathbf{1}$ for odd $N$ and $\text{Hol}(A)=\pm\mathbf{1}$ for even $N$ \cite{Castro:2011iw}, we have
\begin{align}
    a_\tau=-\sum_{i=1}^{\lfloor\frac{N}{2}\rfloor}\tilde{n}_iB^{(1)}_{2i-1}(1,1),
\end{align}
where 
\begin{align}
    \left[B^{(l)}_k(x,y)\right]_{ij}\equiv x\delta_{i,k}\delta_{j,k+l}-y\delta_{i,k+l}\delta_{j,k}.
\end{align}
Here $\tilde{n}_i\in \mathbb{Z}$ for odd $N$ and $\tilde{n}_i\in\mathbb{Z}+1/2$ for even $N$. 
Next, by solving the second EOM in \eqref{BFEOM} for this connection, we obtain 
\begin{align}
    b=h^{-1}\left(-\sum_{i=1}^{\lfloor\frac{N}{2}\rfloor}n_iB^{(1)}_{2i-1}(1,1)\right)h,
\end{align}
where $n_i\in \mathbb{Z}$ for odd $N$ and $n_i+1/2\in\mathbb{Z}$ for even $N$.
These solutions are regarded as the dimensional reduction of solutions in Chern-Simons theory given in \cite{Hikida:2022ltr}. In terms of metric, those solutions correspond to $S^3$ with conical defects. The periodicities for $\tau,\phi$ are given by 
\begin{align}
    2\pi \left(\sum_{i=1}^{\lfloor\frac{N}{2}\rfloor}\frac{2n_i^2}{\epsilon_N}\right)^{-1},\qquad 2\pi \left(\sum_{i=1}^{\lfloor\frac{N}{2}\rfloor}\frac{2\tilde{n}_i^2}{\epsilon_N}\right)^{-1},
\end{align}
respectively.
To impose the spacial periodicity $\phi\sim\phi+2\pi$, we choose $\tilde{n}_i=\rho_i$ with the Weyl vector
\begin{align}
    \rho_i=\frac{N+1}{2}-i.
\end{align}
Compared to \eqref{metric}, the deficit angle is encoded to $r_0$ as
\begin{align}
    r_0=\sum_{i=1}^{\lfloor\frac{N}{2}\rfloor}\frac{2n_i^2}{\epsilon_N}.
\end{align}

Let us evaluate the on-shell action for these solutions.
The bulk action does not again contribute since $f=0$. As the boundary term, here we consider 
\begin{align}
    I_{\text{bdy}}=k\int_{r=r_0} d\tau \tr[b^2],
\end{align}
which is the same one as that given in \cite{Iliesiu:2019xuh}.
This boundary term is actually identical with the one we considered in \eqref{su2bf} for $SU(2)$ BF theory. For this boundary term we find
\begin{align}
    I^{\text{on-shell}}_{\text{G}}=-\frac{\pi r_0}{4G_N},
\end{align}
the twice of which matches with the on-shell action calculated in the second order formalism. 


\section{Dual 2d CFT}
\label{sec:wzw}
In this section, we discuss the reduction of the Chern-Simons formulation of 3d gravity theory to 2d CFT. The procedure is analogous to the ordinary AdS$_3$ case: first reduce two copies of Chern-Simons theory to a non-chiral WZW model and then perform a reduction to the Alekseev-Shatashvili theory, following the discussion of \cite{Cotler:2018zff, Cotler:2019nbi}. 

Here we would like to make a comment to clarify a confusing point about the dual CFT$_2$. In some calculations in \cite{Hikida:2022ltr}, Liouville theory (and Toda field theory) is used instead of $SU(2)_{\hat{k}}$ WZW model. However, we did not use the Hamiltonian reduction to justify using the Liouville theory, but just relied on the fact that the correlators of the Virasoro minimal model can be evaluated by an analytic continuation of Liouville theory \cite{Creutzig:2021ykz}. Therefore this theory is not directly related to the Liouville theory obtained from the Hamiltonian reduction \cite{Coussaert:1995zp}, though there would be some relation between them. Indeed, the Liouville theory used in \cite{Hikida:2022ltr} describes the gravity theory with including the matter sector, while the discussion of Hamiltonian reduction only accounts for the pure gravity sector.

\subsection{Wess-Zumino-Witten model}

First we consider the reduction of the Chern-Simons theory to the Wess-Zumino-Witten model as in the AdS$_3$ case \cite{Witten:1988hf, Moore:1989yh}.
For the Chern-Simons action
	\begin{align}
		I_{\textrm{CS}}[A] \, = \, \frac{k}{4\pi} \int_{M_3} \, \tr \left[ A \wedge dA \, + \, \frac{2}{3} \, A \wedge A \wedge A \right] \, ,
	\end{align}
the equation of motion $F= dA+A \wedge A = 0$ is solved by a pure gauge. Denoting this solution by
\footnote{There could be a zero mode contribution and also a problem about global structure of this solution \cite{Balog:1997zz}. Here, we ignore both of these subtleties.}
	\begin{align}
		A^{\pm} \, = \, G_{\pm}^{-1} d G_{\pm} \, .
	\end{align}
Here we use the notation of \cite{Donnay:2016iyk} to denote $G_{\pm} = G_{\pm}(\t, r, \vp)$ and $g_{\pm} = g_{\pm}(\t, \vp)$.
For the bulk action, we have
	\begin{align}
		I_{\textrm{CS}}[A^{\pm}] \, = \, - \frac{k}{12\pi} \int_{M_3} \, \tr \big( G_{\pm}^{-1} d G_{\pm} \big)^3 \, .
	\end{align}
The original boundary action leads to a coupling between the holomorphic and anti-holomorphic sectors and it is not appropriate.
On the other hand, the modified boundary term gives
	\begin{align}
		\wtd{I}^{(r=r_0)}_{\textrm{bdy}}[A^+, A^-]
        \, = \, \frac{k}{2\pi} \int_{r=r_0} d\t d\vp \tr \Big[ - g_+^{-1} \pa_\t g_+ g_+^{-1} \pa_+ g_+ \, + \, g_-^{-1} \pa_\t g_- g_-^{-1} \pa_- g_- \Big] \, ,
	\end{align}
where $\pa_{\pm} = \frac{1}{2}(\pa_\t \pm \pa_\vp)$.
This certainly leads to a decoupled action between the holomorphic and anti-holomorphic sector.
Therefore, the total CFT action is given by
	\begin{align}
		I_{\textrm{CFT}}[g_+, g_-] \, = \, I_{\textrm{WZW}}^{(+)}[g_+] \, - \, I_{\textrm{WZW}}^{(-)}[g_-] \, ,
    \label{I_CFT}
    \end{align}
with
	\begin{align}
		I_{\textrm{WZW}}^{(\pm)}[g_\pm] \, = \, \mp \frac{k}{2\pi} \int_{\pa M_3} d\t d\vp \tr \big( g_\pm^{-1} \pa_\t g_\pm g_\pm^{-1} \pa_\pm g_\pm \big) \Big|_{r=r_0}
        - \, \frac{k}{12\pi} \int_{M_3} \tr \big( G_\pm^{-1} d G_\pm \big)^3 \, .
    \end{align}
This CFT is indeed the one considered in \cite{Hikida:2021ese, Hikida:2022ltr}, which was purposed in to be dual to the 3D gravity on $S^3$.

We note that for this WZW model, the level $k$ is defined in (\ref{level-k}), and the classical limit corresponds to $k \to \inf$.
The on-shell action agrees with the one obtained in the second order formalism in this limit as we showed in section~\ref{sec:3d gravity}.
On the other hand, the WZW model used in \cite{Hikida:2021ese, Hikida:2022ltr} has level $\hat{k}$ and it was claimed that the $\hat{k} \to -2 + 4G_N i$ limit is dual to the 3D gravity.
We will clarify some of the relations between these two limits, but it is interesting to understand more the connection between these two limits of the different WZW models.

\subsection{Alekseev-Shatashvili theory}
\label{sec:ak}
The reduction of the WZW model to Liouville theory is well-known at least for $SL(2,\mathbb{R})$ case \cite{Donnay:2016iyk, Forgacs:1989ac, Alekseev:1988ce}.
However, as pointed out in \cite{Cotler:2018zff, Cotler:2019nbi}, this reduction is not complete and one needs to perform a quasi-local quotient.
This procedure leads to a two copies of the Alekseev-Shatashvili action.
In this subsection, we assume the simple analytic continuation from AdS$_3$ to $S^3$ (as we reviewed in appendix~\ref{app:rotation})
also works for this rewriting of the WZW model to the Alekseev-Shatashvili action for our $SU(2)$ case.

Following \cite{Cotler:2018zff, Cotler:2019nbi}, we perform a Gauss decomposition of the gauge field $G_{\pm}$ on the $SU(2)$ basis (\ref{SU(2)_generators}):
	\begin{align}
		G_+ \, 
		&= \, \Bigg( \begin{matrix} \cos\left( \frac{X}{2} \right) & -\sin\left( \frac{X}{2} \right) \\[4pt]
        \sin\left( \frac{X}{2} \right) & \cos\left( \frac{X}{2} \right) \end{matrix} \Bigg)
        \Bigg( \begin{matrix} \L & 0 \\ 0 & \L^{-1} \end{matrix} \Bigg)
        \Bigg( \begin{matrix} 1 & Y \\ 0 & 1 \end{matrix} \Bigg) \, .
	\end{align}
Here, we note that the fields we introduced here are all complex fields with Re$(\L)>0$ and $X \sim X + 2\pi$.
One can think that this process is extending the original $SU(2)$ gauge symmetry to the $SL(2; \mathbb{C})$ symmetry at this point.
For $G_-$, we have the same form of decomposition, but with fields $\bar{X}$, $\bar{\L}$ and $\bar{Y}$.
If we require the $G_+$ to be unitary, the fields must to satisfy the following conditions.
    \begin{align}
        |\Lambda|^2\cos\left(\frac{X-X^*}{2}\right)\, = \, 1 \, , \qquad
        Y=\frac{\Lambda^*}{\Lambda}\sin\left(\frac{X-X^*}{2}\right)
    \end{align}
After substituting this decomposition into the action and then imposing the quasi-local quotient leads to the Alekseev-Shatashvili action \cite{Cotler:2018zff, Cotler:2019nbi}
	\begin{align}
		I_{\textrm{CFT}}[X, \bar{X}] \, = \, I_{\textrm{AS}}^{(+)}[X] \, + \, I_{\textrm{AS}}^{(-)}[\bar{X}] \, ,
    \end{align}
with
	\begin{align}
		I_{\textrm{AS}}^{(\pm)}[X] \, = \, \frac{k}{\pi} \int d\t d\vp \left[ \frac{X''\pa_{\pm} X'}{X'^2} \, - \, \m \, X' \pa_{\pm} X \right] \, ,
    \label{I_AS}
    \end{align}
where the prime denotes a derivative with respect to $\t$ and $\m$ is a complex constant.

The holomorphic part of the stress tensor is given by
	\begin{align}
		T \, = \, - \frac{c}{12} \ S\Big( \tan(\tfrac{X}{2}); x^+ \Big) \, ,
    \end{align}
where $S\big( f; z \big)$ is the Schwarzian derivative
	\begin{align}
		S\big( f; z \big) \, = \, \frac{f'''(z)}{f''(z)} \, - \, \frac{3}{2} \left( \frac{f''(z)}{f'(z)} \right)^2 \, ,
    \end{align}
and the central charge is
	\begin{align}
		c \, = \, 6k \, .
    \label{c_v}
    \end{align}

\section{Liouville quantum mechanics}
\label{sec:liouville-qm}

\subsection{From Alekseev-Shatashvili theory}

Finally, we consider the dimensional reduction of the Alekseev-Shatashvili action defined in (\ref{I_AS}).
As in section~\ref{sec:3d gravity} and~\ref{sec:bf}, we regard the field $X$ (and $\bar{X}$) independent of the spacial coordinate $\vp$.
This leads to 
	\begin{align}
		I_{\textrm{AS}}^{(+)}[X] \, = \, 2k \int_0^\b d\t \left[ \frac{X''^2}{X'^2} \, - \, \m \, X'^2 \right] \, ,
    \end{align}
If we introduce the Liouville variable by $X' = e^{\frac{i \p}{2}}$, then we arrive at the Liouville quantum mechanics
	\begin{align}
		I_{\textrm{LQM}}[\p] \, = \, I_{\textrm{AS}}^{(+)}[X] \, = \, - k \int_0^\b d\t \left[ \frac{1}{2} \, ( \pa_\t \p )^2 + \, 2 \m \, e^{i\p} \right] \, .
    \label{I_LQM}
    \end{align}
Since the conjugate momentum is $\pi_\p = \frac{\pa L}{\pa \dot{\p}}=-k \dot{\p}$, where the dot denotes a derivative with respect to $\t$, the Lagrangian is written as
	\begin{align}
		L \, = \, \pi_\p \dot{\p} \, + \, \frac{\pi_\p^2}{2k} \, - \, 2 k\m \, e^{i\p} \, .
    \end{align}
In (\ref{I_AS}), we took a particular solution where $\m$ is a constant.
Now we imagine to promote the Liouville cosmological constant $\m$ to be a time-dependent variable.
Therefore, the action (\ref{I_LQM}) is a particular gauge fixed version of this generalized theory.
In order to recover the gauge degrees of freedom, we promote the Liouville cosmological constant $\m$ to a dynamical valuable by $\pi_f= -4 i k^2 \m$,
where the coefficient is just for convenience.
Adding the kinetic term for this degrees of freedom, we have new Lagrangian
	\begin{align}
		L \, = \, \pi_\p \dot{\p} \, + \, \pi_f \dot{f} \, + \, \frac{1}{2k} \Big( \pi_\p^2 \, - \, i \pi_f \, e^{i \p} \Big) \, .
    \label{L_LQM}
    \end{align}
Therefore, the Hamiltonian is $H=\frac{1}{2k}(-\pi_\p^2 + i \pi_f e^{i\p})$.

This theory now has the $SU(2)$ gauge symmetry, which is generated by
	\begin{align}
		&\hat{\ell}_{-1} \, = \, i \pi_f \, , \\
		&\hat{\ell}_{0} \, = \, - f \pi_f \, + \, i \pi_\p \, , \\
		&\hat{\ell}_{+1} \, = \, i f^2 \pi_f + 2f \pi_\p  + e^{i\p} \, .
    \label{su(2)_generators}
    \end{align}
Imposing the canonical commutation relations $[\p, \pi_\p] = 1 = [f, \pi_f]$, these generators satisfy
	\begin{align}
		[\hat{\ell}_{\pm1}, \hat{\ell}_0] \, = \, \pm \, \hat{\ell}_{\pm1} \, , \qquad
        [\hat{\ell}_{+1}, \hat{\ell}_{-1}] \, = \, 2 \hat{\ell}_{0} \, .
    \end{align}
The Hamiltonian is now given by the $SU(2)$ quadratic Casimir as
	\begin{align}
		H \, = \, \frac{1}{2k} \left( \hat{\ell}_{0}^{\, 2} + \frac{1}{2} \{\hat{\ell}_{-1}, \hat{\ell}_{+1} \} \right)
        \, = \, \frac{1}{2k} \Big( - \pi_\p^2 \, + \, i \pi_f \, e^{i\p} \Big) \, .
    \label{Hamiltonian}
    \end{align}

Next we consider the partition function of this theory in the spin $j$ representation.
Since the Hamiltonian is the $SU(2)$ quadratic Casimir, the partition function can be written in terms of the $SU(2)$ characters as
	\begin{align}
		Z_j(\b) \, = \, \Tr_{R_j} \big[ e^{- \b H } \big] \, = \, \c_j\left(\frac{i\b}{4\pi k} \right) \, ,
    \end{align}
where the character is
	\begin{align}
		\c_j(\t) \, = \, \Tr_{R_j} \big[ e^{2\pi i \t H } \big] \, .
    \end{align}
Since we are interested in the semi-classical limit $k \to \infty$, it is more useful to consider the $S$-modular transform of the character \cite{Mertens:2017mtv, Ghosh:2019rcj}.
Under the $S$-transform $\t \to - 1/\t$, the character transforms as
	\begin{align}
		\c_j(\t) \, = \, \sum_\ell S_j{}^\ell \, \c_\ell(-1/\t) \, ,
    \label{S-transf}
    \end{align}
with the $S$ matrix
	\begin{align}
		S_j{}^\ell \, = \, \sqrt{\frac{2}{\hat{k}+2}} \, \sin\left( \frac{\pi}{\hat{k}+2} (j+1)(\ell+1) \right) \, .
    \end{align}
where
    \begin{align}
		\hat{k} \, = \, \frac{4k}{1-2k} \, .
    \end{align}  
Since
	\begin{align}
		\c_j(-1/\t) \, = \, \c_j\left( \frac{4\pi k i}{\b} \right) \, = \, \Tr_{R_j} \big[ e^{- \frac{8\pi^2 k H}{\b}} \big] \, ,
    \end{align}
the ground state dominates among the summation in (\ref{S-transf}) in the semi-classical limit $k \to \inf$.
Therefore, in the semi-classical limit, the partition function is approximated by
    \begin{align}
		Z_j(\b) \, \approx \, S_j{}^0 \, = \, \sqrt{\frac{2}{\hat{k}+2}} \, \sin\left( \frac{\pi r_0}{\hat{k}+2} \right) \, ,
    \label{Z_j}
    \end{align}
where we used $r_0 = j+1$. A similar discussion was also given in \cite{Mertens:2018fds}.
The rest of the story is as same as in the higher dimensional case proposed in \cite{Hikida:2021ese, Hikida:2022ltr}.
In the 
    \begin{align}
		\hat{k} \, \to \, -2 \, + \, \frac{6i}{c^{(g)}} \, + \, \mathcal{O}\big( (c^{(g)})^2 \big) \, ,
    \label{k-limit}
    \end{align}
limit, where
    \begin{align}
		c^{(g)} \, = \, \frac{3}{2G_N} \, ,
    \end{align}
the partition function becomes
    \begin{align}
		Z_j(\b) \, \approx \, \left| \frac{c^{(g)}}{3} \right|^{\frac{1}{2}} \, \exp\left[ \frac{\pi c^{(g)} r_0}{6} \right] \, .
    \end{align}
Therefore, square of this partition function agrees with the JT gravity result in (\ref{I_onshell_2D}).

We note that the level $k:=k_{\textrm{CS}}$ in AS action is the Chern-Simons level and it is different from the level $\hat{k}$ of the $SU(2)_{\hat{k}}$ WZW model considered in \cite{Hikida:2021ese, Hikida:2022ltr}.
In the $SU(2)_{\hat{k}}$ WZW model, the central charge is expressed by
	\begin{align}
		c \, = \, \frac{3\hat{k}}{\hat{k}+2} \, .
    \end{align}
If we equate this expression to (\ref{c_v}) and solve it for $\hat{k}$, we find the relation between the two levels
    \begin{align}
		\hat{k} \, = \, \frac{4k_{\textrm{CS}}}{1-2k_{\textrm{CS}}} \, .
    \end{align}  
From this equation, we can see that the semi-classical limit of the Chern-Simons level corresponds to the critical limit of the WZW level:
	\begin{align}
		k_{\textrm{CS}} \to \infty \qquad \Leftrightarrow \qquad \hat{k} \to -2 \, .
    \end{align}
This is the origin of the critical limit of the WZW level proposed in \cite{Hikida:2021ese,Hikida:2022ltr}.

We can also recover the above result for the partition function by studying the LQM in the semi-classical limit (in large $k=k_{\textrm{CS}}$).
The equation of motion of the action (\ref{I_LQM}) is 
    \begin{align}
		-\pa_\t^2 \phi \, + \, 2i \m \, e^{i \phi} \, = \, 0 \, ,
    \end{align}
and the periodic boundary condition $\phi(\t)=\phi(\t+\b)$ gives the background solution
    \begin{align}
		e^{i \phi(\t)} \, = \, \frac{1}{\m} \left( \frac{\pi}{\b} \right)^2 \frac{1}{\cos\big(\frac{\pi}{\b} (\t+c) \big)^2} \, ,
    \end{align}
where $c$ is a real constant. Once we plug this solution back into the action (\ref{I_LQM}), we find
	\begin{align}
		I_{\textrm{LQM}}^{\onshell} \, = \, - \frac{2\pi^2k}{\b} \, = \, - \frac{\pi^2}{\b} \frac{\hat{k}}{\hat{k}+2}
  \, \approx \, - \frac{\pi^2 c^{(g)}}{3\b} \, = \, - \frac{\pi c^{(g)} r_0}{6} \, .
    \label{I_LQM^onshell}
	\end{align}
This agrees with (\ref{Z_j}).

Based on this observation, we finally propose our conjecture that the JT gravity defined on $S^2$ as in (\ref{I_JT}) is dual to 
two copies of the Liouville quantum mechanics defined in this section in the critical limit of $\hat{k}$ (\ref{k-limit}).

\subsection{From Schwarzian theory}
In this subsection, we discuss how to derive the Liouville quantum mechanics we obtained in the previous subsection from the Schwarzian theory.
The Schwarzian theory is described by the Lagrangian
    \begin{align}
		L \, = \, C \, S(f; \t) \, = \, C \left[ \frac{f'''(\t)}{f''(\t)} \, - \, \frac{3}{2} \left( \frac{f''(\t)}{f'(\t)} \right)^2 \right] \, ,
    \end{align}
where $C$ is the Schwarzian coupling and now the prime denotes the derivative with respect to $\t$.
This theory has $SL(2; \mathbb{R})$ symmetry, but now we complexify the field $f$ and the time $\t$.
Now the theory is invariant under the $SL(2; \mathbb{C})$ symmetry
    \begin{align}
		f(\t) \, \to \, \frac{\a f(\t) + \b}{\g f(\t) + \d} \, , \qquad \quad (\a, \b, \g, \d \, \in \, \mathbb{C})
    \end{align}
If we choose a particular direction of this $SL(2; \mathbb{C})$ symmetry, we find the (real) $SU(2)$ symmetry.
The infinitesimal transformations of the $SU(2)$ symmetry is given by
    \begin{align}
		\d_- f(\t) \, &= \, i \ve_- \, , \\
		\d_0 \, f(\t) \, &= \, - \ve_0 \, f(\t) \, , \\
		\d_+ f(\t) \, &= \, i \ve_+ \, f(\t)^2 \, ,
    \end{align}
where $\ve_0, \ve_-, \ve_+ \in \mathbb{R}$.
The associated conserved charges of these transformations are found as \cite{Maldacena:2016upp}
    \begin{align}
		Q_- \, &= \, i C \left[ \frac{f'''}{f'^2} \, - \, \frac{f''^2}{f'^3} \right] \, , \\
		Q_0 \, &= \, - C \left[ \frac{ff'''}{f'^2} \, - \, \frac{ff''^2}{f'^3} \, - \, \frac{f''}{f'} \right] \, , \\
		Q_+ \, &= \, i C \left[ \frac{f^2f'''}{f'^2} \, - \, \frac{f^2f''^2}{f'^3} \, - \, \frac{2ff''}{f'} \, + \, 2f' \right] \, .
    \end{align}
With appropriate canonical quantization, these charges satisfy
    \begin{align}
		[Q_0 , Q_\pm] \, = \, \pm Q_\pm \, , \qquad [Q_+, Q_-] \, = \, 2 Q_0 \, .
    \label{su(2)_commutations}
    \end{align}

In order to see this canonical quantization, we rewrite the Lagrangian by integration by parts as 
    \begin{align}
		L \, = \, \frac{C}{2} \, \p'^2 \, + \, \pi_f (f' - i e^{i \p}) \, ,
    \end{align}
where $\pi_f$ is the Lagrange multiplier and we used an identification $f' = i e^{i\p}$ \cite{Mertens:2017mtv}.
Since $\pi_\p = C \p'$, we can also rewrite this Lagrangian as
    \begin{align}
		L \, = \, \pi_\p \p' \, + \, \pi_f f' - \left( \frac{\pi_\p^2}{2C} \, + \, i \pi_f e^{i \p} \right) \, .
    \end{align}
The equation of motion for $\p$ gives us $C\p'' = \pi_f e^{i\p}$.
Using this equation and the identification $f' = i e^{i\p}$, we can rewrite the conserved charges as
    \begin{align}
		Q_- \, &= \, i \pi_f \, , \\
		Q_0 \, &= \, - f \pi_f \, + \, i \pi_\p \, , \\
		Q_+ \, &= \, i f^2 \pi_f \, + \, 2 f\pi_\p \, - \, 2 C e^{i\p} \, .
    \end{align}
If we impose the canonical quantization $[\p, \pi_\p] = [f, \pi_f] = 1$, we find the charges satisfy the $SU(2)$ commutation relations (\ref{su(2)_commutations}).
The Hamiltonian is now given by
	\begin{align}
		H \, = \, \frac{1}{2C} \left( Q_{0}^{\, 2} + \frac{1}{2} \{Q_-, Q_+ \} \right)
        \, = \, \frac{1}{2C} \Big( - \pi_\p^2 \, - \, 2iC \pi_f \, e^{i\p} \Big) \, .
    \end{align}
Now these charges and the Hamiltonian agree with (\ref{su(2)_generators}) and (\ref{Hamiltonian}),
once we rescale $f \to -2C f$ and $\pi_f \to - \pi_f/2C$ with identifications $Q_-=-\hat{\ell}_{-1}/2C$, $Q_0=\hat{\ell}_0$ and $Q_+=-2C \hat{\ell}_{+1}$.

\section{Conclusions and discussions}
\label{sec:conclusions}
In this paper, we have studied a dimensional reduction of the $S^3$/WZW model duality proposed in \cite{Hikida:2021ese, Hikida:2022ltr}.
Justification of this dimensional reduction comes from the fact that the classical on-shell action in the duality of \cite{Hikida:2021ese, Hikida:2022ltr} is simply linear in temperature. 
For the gravity side, in the second order formalism, we showed that the Einstein gravity on $S^3$ is simply reduced to a JT gravity on $S^2$ with a non-standard (or Neumann-Neumann type) boundary term.
In the first order formalism, we have discussed an inconsistency between the boundary term, which is simply obtained by rewriting from the second order formalism action, and the background solution. 
We then provided a possible modification of the boundary term,
even though this may not be the unique modification of the boundary term which is consistent with the background solutions.
For the CFT side we argued that the dimensional reduction of the $SU(2)_{\hat{k}}$ WZW model is given by a certain type of a complex Liouviile quantum mechanics with $SU(2)$ gauge symmetry. therefore, we proposed a duality between the $S^2$ (classical) JT gravity and this complex Liouville quantum mechanics in the limit of $\hat{k} \to -2$.

There are several future directions from this work.
One obvious direction is to study correlation functions in our proposed duality which is work in progress \cite{Suzuki:2023}.
In particular, it is interesting to study the out-of-time order correlators in this duality.
In the AdS$_2$ JT gravity/the Schwarzian theory duality \cite{Maldacena:2019cbz}, the out-of-time order correlators saturates the known maximal chaos bound \cite{Maldacena:2015waa}.
It would be interesting to check weather we have the same story in our $S^2$ JT gravity/complex Liouville quantum mechanics duality.

Another direction would be to study a supersymmetric generalization of our works.
Since the supersymmetric WZW model \cite{DiVecchia:1984nyg, Kac:1985wdn, Fuchs:1986ew} has a similar structure of the central charge and $S$-matrix, it seems possible to consider the smae limit proposed in \cite{Hikida:2021ese, Hikida:2022ltr} for this case as well. 
It would be interesting to explicitly check whether this limit produces the corresponding three-dimensional supergravity.
If this works, it is also interesting to study its dimensional reduction to see if it produces a supersymmetric JT gravity and supersymmetric Liouville quantum mechanics.

\section*{Acknowledgements}
We are grateful to Yasuaki Hikida and Tadashi Takayanagi for useful comments on the draft of this paper.
This work is supported by MEXT KAKENHI Grant-in-Aid for Transformative Research Areas (A) through the ``Extreme Universe'' collaboration: Grant Number 21H05187. 
The work of KS is also supported by JSPS KAKENHI Grant No. 23K13105.
The work of YT is supported by  Grant-in-Aid for JSPS Fellows No. 22KJ1971.

\appendix
\section{$SU(2)$ symmetry}
\label{app:su(2)}
In this appendix, we summarize our notation for the $SU(2)$ symmetry.
We define the $SU(2)$ generators 
    \begin{align}
		J_1 \, = \, \frac{i \s_2}{2} \, = \, \frac{1}{2} \begin{pmatrix} 0 & -1 \\ 1 & 0 \end{pmatrix} \, , \ \
		J_2 \, = \, \frac{i \s_3}{2} \, = \, \frac{1}{2} \begin{pmatrix} i & 0 \\ 0 & -i \end{pmatrix} \, , \ \
		J_3 \, = \, \frac{i \s_1}{2} \, = \, \frac{1}{2} \begin{pmatrix} 0 & i \\ i & 0 \end{pmatrix} \, .
	\label{SU(2)_generators}
	\end{align}
which satisfy
	\begin{align}
		\big[ J_A, J_B \big] \, = \, \e_{ABC} J^C \, , \qquad \tr\big( J_A J_B \big) \, = \, - \frac{1}{2} \, \d_{AB} \, , \qquad \tr\big( J_A J_B J_C \big) \, = \, - \frac{1}{4} \, \e_{ABC} \, .
	\end{align}
Sometimes it's also useful to define the Chevalley basis by $\L_0 = i J_2$, $\L_\pm = - i(J_3 \pm i J_1)$, which satisfy
    \begin{align}
		\big[ \L_0, \L_\pm \big] \, = \, \pm \L_\pm \, , \qquad \big[ \L_+, \L_- \big] \, = \, 2 \L_0 \, .
	\end{align}

Let us also summarize the relation of these $SU(2)$ generators to the $SL(2;\mathbb{R})$ generators.
We denote that latter generators with tilde. They can be introduced by $\tilde{J}_1=i J_1$, $\tilde{J}_2= - J_2$, $\tilde{J}_3=i J_3$ and they satisfy
    \begin{align}
		\big[ \tilde{J}_1, \tilde{J}_2 \big] \, = \, - \, \tilde{J}_3 \, , \quad
		\big[ \tilde{J}_2, \tilde{J}_3 \big] \, = \, - \, \tilde{J}_1 \, , \quad
		\big[ \tilde{J}_3, \tilde{J}_1 \big] \, = \, \tilde{J}_2 \, .
	\end{align}
Their Chevalley basis are given by $\tilde{\L}_0 = i \tilde{J}_2=-\L_0$, $\tilde{\L}_\pm = - i(\tilde{J}_3 \pm i \tilde{J}_1)=i \L_\pm$, which satisfy
    \begin{align}
		\big[ \tilde{\L}_0, \tilde{\L}_\pm \big] \, = \, \mp \tilde{\L}_\pm \, , \qquad \big[ \tilde{\L}_+, \tilde{\L}_- \big] \, = \, 2 \tilde{\L}_0 \, .
	\end{align}

\section{Boundary conditions of JT gravity}
\label{app:bc_jt}
In this appendix, we discuss the details of the variational principle and boundary conditions for the JT gravity, we derived in section~\ref{sec:jt}, mostly following the discussion of \cite{Goel:2020yxl}.
By taking the variation of the bulk JT action
	\begin{align}
		I_{\textrm{bulk}} \, = \, - \frac{1}{8 G_N} \int d^2x \sqrt{g_2} \, \P \big( R_2 - 2 \big) \, ,
	\end{align}
we find \cite{Goel:2020yxl}
	\begin{align}
		\d I_{\textrm{bulk}} \, = \, - \textrm{EOM} \, + \, \frac{1}{8G_N} \int du \Big[ 2 \P \d\big( \sqrt{g_{uu}} \, K \big) - 2 \pa_n \P \big( \d \sqrt{g_{uu}} \big) \Big] \, ,
	\end{align}
where $u$ is the boundary Euclidean time and $K$ is the trace of the extrinsic curvature on the boundary.
Therefore, with the boundary term in (\ref{I_JT}), we have boundary conditions
	\begin{align}
		\sqrt{g_{uu}} \, K \, , \qquad \textrm{and} \qquad \pa_n \P \qquad \textrm{fixed}
	\end{align}
where $\pa_n$ is the normal derivative with respect to the boundary.

In order to parametrize the boundary value of $K$, let us consider the Fefferman-Graham-like coordinates
	\begin{align}
		ds^2 \, = \, \frac{\big(\sqrt{r_0^2-r^2} - b(\t) r\big)^2}{1+b(\t)^2} \, d\t^2 \, + \, \frac{dr^2}{r_0^2 - r^2} \, ,
    \label{FG-like}
    \end{align}
where $b(\t)$ is an arbitrary (smooth) function of $\t$.
This metric gives $R_2 = 2$, regardless of the exact form of $b(\t)$. 
This metric can be realized in the three-dimensional embedding space
	\begin{align}
		X^2 \, + \, Y^2 \, + \, Z^2 \, = \, 1 \, ,
	\end{align}
with parametrizations
	\begin{align}
		X \, &= \, \Bigg( \sqrt{1- \left( \frac{r}{r_0} \right)^2} - b(\t) \frac{r}{r_0} \Bigg) \frac{\cos(r_0 \t)}{\sqrt{1+b(\t)^2}} \, , \\
		Y \, &= \, \Bigg( \sqrt{1- \left( \frac{r}{r_0} \right)^2} - b(\t) \frac{r}{r_0} \Bigg) \frac{\sin(r_0 \t)}{\sqrt{1+b(\t)^2}} \, , \\
		Z \, &= \, \Bigg( \frac{r}{r_0} + b(\t) \sqrt{1- \left( \frac{r}{r_0} \right)^2} \Bigg) \frac{1}{\sqrt{1+b(\t)^2}} \, .
	\label{embedding}
    \end{align}
With this metric, the trace of the extrinsic curvature is given by
	\begin{align}
		K \, = \, \frac{1}{b(\t)} \left( \frac{r + b(\t) \sqrt{r_0^2 - r^2}}{r - b(\t)^{-1} \sqrt{r_0^2 - r^2}} \right) \, .
	\end{align}
Therefore the first boundary condition which fixes $\sqrt{g_{uu}} K$ on the boundary, fixes $b(\t) = b_0 =$ constant.

Now under the metric (\ref{FG-like}) with $b(\t) = b_0$, the dilaton equation (\ref{dilaton-eq}) is solved by 
	\begin{align}
        \P \, = \, \Big[ A X + B Y + C Z \Big]_{b(\t) = b_0} \, ,
	\end{align}
where $X, Y, Z$ are given by (\ref{embedding}) and $A, B, C$ are integration constants. 
Now the second boundary condition for the normal derivative of the dilaton 
	\begin{align}
        \pa_n \P \, =  \, \sqrt{r_0^2 - r^2} \, \pa_r \P \, = \, \textrm{fixed}
	\end{align}
requires $A=B=0$. Hence, the values of $b_0$ and $C$ finally controls all possible boundary conditions.
The solution we discussed in the main text corresponds to $b_0 = 0$ and $C=r_0$.

\section{Wick rotation from AdS$_3$}
\label{app:rotation}
In this appendix, let us review the Wick rotation from (Lorentzian) AdS$_3$ to $S^3$.
AdS$_3$ can be embedded into $\mathbb{R}^{2,2}$ as a timelike hypersurface as
	\begin{align}
		- \big( X^{-1} \big)^2 - \big( X^0 \big)^2 + \big( X^1 \big)^2 + \big( X^2 \big)^2 \, = \, - \, L_{\textrm{AdS}}^2 \, .
	\end{align}
The global coordinates of AdS$_3$ is implemented by
	\begin{align}
		X^{-1} \, &= \, L_{\textrm{AdS}} \, \cosh \r \, \cos t \, , \\
		X^0 \, &= \, L_{\textrm{AdS}} \, \cosh \r \, \sin t \, , \\
		X^1 \, &= \, L_{\textrm{AdS}} \, \sinh \r \, \cos \vp \, , \\
		X^2 \, &= \, L_{\textrm{AdS}} \, \sinh \r \, \sin \vp \, ,
	\end{align}
which leads to 
	\begin{align}
		ds^2 \, = \, L_{\textrm{AdS}}^2 \big( - \cosh^2 \r \, dt^2 + d\r^2 + \sinh^2 \r \, d\vp^2 \big) \, .
	\end{align}

Now the double Wick Rotation to $S^3$ is obtained by
	\begin{align}
		L_{\textrm{AdS}} \, \to \, i L_{S^3} \, , \qquad \r \, \to \, i \th \, .
	\end{align}
This transforms the metric to 
	\begin{align}
		ds^2 \, = \, L_{S^3}^2 \big( d\th^2 + \cos^2 \th \, dt^2 + \sin^2 \th \, d\vp^2 \big) \, .
	\end{align}
We note that $t$ was originally a Lorentzian time in AdS$_3$, but now it becomes an Euclidean time in $S^3$ without any Wick Rotation for it. 

If we recover $L_{S^3}$ in (\ref{metric}), we have 
	\begin{align}
		E_M{}^A \, \propto \, L_{S^3} \, , \qquad \O_M{}^A \, \propto \, 1 \, .
	\end{align}
This explains why there is no explicit imaginary unit appears in the definition of the gauge fields in (\ref{gaugeA}),
where the imaginary unit from $\sqrt{\L}$ cancels with the one from $E_M{}^A$.

For the discussion in section~\ref{sec:ak}, it is also useful to have the Wick rotation from the (Lorentzian) BTZ black hole metric to the Eulidean dS black hole metric.
The (Lorentzian) BTZ black hole metric is 
	\begin{align}
		ds^2 \, = \, L_{\textrm{AdS}}^2 \left( - (r^2 - r_0^2) dt^2 \, + \, \frac{dr^2}{r^2 - r_0^2} \, + \, r^2 d\th^2 \right) \, .
	\end{align}
The Wick rotation to the Eulidean dS black hole metric is given by
	\begin{align}
		L_{\textrm{AdS}} \, \to \, i L_{S^3} \, , \qquad r \, \to \, i r \, \qquad (r_0 \, \to \, i r_0 \, , \qquad t \, \to \t ) \, .
	\end{align}
This leads to the Eulidean dS black hole metric
	\begin{align}
		ds^2 \, = \, L_{S^3}^2 \left( (r_0^2 - r^2) d\t^2 \, + \, \frac{dr^2}{r_0^2 - r^2} \, + \, r^2 d\th^2 \right) \, .
	\end{align}

\section{Chern-Simons gravity on $B^3$}\label{CSonB3}

As discussed in the introduction, the partition function on $S^3$ is constructed as the squared norm of the Hartle-Hawking wave functional of universe and then a double copy of the CFT partition functions:
\begin{align}
    Z_{\text{G}}[S^3]=|\Psi_{\text{HH}}|^2=|Z_{\text{CFT}}[S^2]|^2.
\end{align}
In the semi-classical limit, the Hartle-Hawking wave function takes the form 
\begin{align}
    \Psi_{\text{HH}}\sim \exp\left(I^{(E)}_{\text{G}}[B^3]+iI^{(L)}_{\text{G}}[\text{dS}_{t>0}]\right).
\end{align}
Since the Lorentzian part is just a phase factor, only the Euclidean part survives when we take the squared norm and gives $S^3$ partition function. We split the sphere as $S^3=B^3\cup \overline{B^3}$ (glued along the equator) and apply CS/WZW correspondence (with boundary) to each $B^3$. In this perspective we expect that a double copy of WZW models located on the equator is dual to the Chern-Simons gravity on $S^3$.

\subsection{Chern-Simons gravity on $B^3$}

Let us consider the Chern-Simons gravity on $B^3$ with unit radius, whose metric is given by
\begin{align}
    ds^2=d\theta^2+\sin^2\theta \,ds_{S^2}^2,\quad 0\le\theta\le\frac{\pi}{2}.
\end{align}
with the boundary at $\theta=\pi/2$.
This metric can be obtained by the analytic continuation $T\to i\theta+i\pi/2$ from Lorentzian dS$_3$ metric
\begin{align}
    ds^2=-dT^2+\cosh^2T\, ds_{S^2}^2.
\end{align}
As the metric of $S^2$, we adopt the Fubini-Study metric for $\mathbb{C}P^1\simeq S^2$:
\begin{align}
    ds_{S^2}^2=\frac{4dzd\bar{z}}{(1+|z|^2)^2}=\frac{4(dx^2+dy^2)}{(1+x^2+y^2)^2},
\end{align}
where $z=x+iy, \bar{z}=x-iy$.
The vielbein and the spin connection associated with this metric are 
\begin{alignat}{3}
    &E^1 =d\theta,& \quad &E^2=\frac{2\sin\theta}{1+x^2+y^2}dx,&\quad &E^3=\frac{2\sin \theta}{1+x^2+y^2}dy\\
    &\Omega^1 = \frac{2(ydx-xdy)}{1+x^2+y^2},&\quad &\Omega^2=-\frac{2\cos\theta}{1+x^2+y^2}dy,&\quad &\Omega^3=\frac{2\cos\theta}{1+x^2+y^2}dx.
\end{alignat}
Then the connections are
\begin{align}\label{FSconnection}
    A^{\pm}=\pm \frac{1}{2}
            \begin{pmatrix}
                i & 0\\
                0 & -i
            \end{pmatrix}
            d\theta+\frac{1}{2(1+|z|^2)}
            \begin{pmatrix}
                zd\bar{z}-\bar{z}dz & -2e^{\mp i\theta}dz \\
                2e^{\pm i\theta}d\bar{z} & -zd\bar{z}+\bar{z}dz
            \end{pmatrix}.
\end{align}
On the boundary $\theta=\pi/2$, each component of these fields takes the following form
\begin{align}\label{bdycomponents}
    A^{\pm}_z=\frac{1}{2(1+|z|^2)}
    \begin{pmatrix}
        -\bar{z} & \pm 2i \\
        0 & \bar{z} 
    \end{pmatrix},\qquad
    A^\pm_{\bar{z}}=\frac{1}{2(1+|z|^2)}
    \begin{pmatrix}
        z & 0 \\
        \pm2i & -z
    \end{pmatrix}.
\end{align}

We would like to consider the boundary term of the Chern-Simons gravity action.
Chern-Simons action is defined by
\begin{align}
    I_{\text{CS}}[A]=\frac{k}{4\pi}\int\Tr\left[A\wedge dA+\frac{2}{3}A\wedge A\wedge A\right].
\end{align}
In terms of the coordinates introduced above, this can be expressed as
\begin{align}
    I_{\text{CS}}[A]=-\frac{k}{4\pi}\int_{B^3}d\theta dxdy\Tr[A_y\partial_\theta A_x-A_x\partial_\theta A_y+2A_\theta F_{xy}]
\end{align}
without producing any boundary contributions. 
The variation of the holomorphic part of the action is
\begin{align}
    \delta I_{\text{CS}}[A]&=(\text{EOM})+\frac{k}{4\pi}\int_{\theta=\pi/2}dxdy\Tr[A_y\delta A_x-A_x\delta A_y]\\
                        &=(\text{EOM})-\frac{k}{4\pi}\int_{\theta=\pi/2} dzd\bar{z}\Tr[A_z\delta A_{\bar{z}}-A_{\bar{z}}\delta A_z].
\end{align}
Now we consider the following boundary term:
\begin{align}
    I_{\text{bdy}}^+[A]=\frac{k}{4\pi}\int_{\theta=\pi/2}dxdy\Tr[A_zA_{\bar{z}}],
\end{align}
then the variation of the total action becomes 
\begin{align}
    \delta(I_{\text{CS}}[A]+I_{\text{bdy}}^+[A])=(\text{EOM})+\frac{k}{2\pi}\int_{\theta=\pi/2}\Tr[A_{\bar{z}}\delta A_z]
\end{align}
Therefore we can fix the boundary value of $A_z$. On the other hand, if we flip the overall sign as 
\begin{align}
    I_{\text{bdy}}^-[A]=-\frac{k}{4\pi}\int_{\theta=\pi/2}dxdy\Tr[A_zA_{\bar{z}}],
\end{align}
then the variation of the total action
\begin{align}
    \delta(I_{\text{CS}}[A]+I_{\text{bdy}}^-[A])=(\text{EOM})-\frac{k}{2\pi}\int_{\theta=\pi/2}\Tr[A_z\delta A_{\bar{z}}],
\end{align}
which fixes $A_{\bar{z}}$ on the boundary. Now our interest is the difference of two Chern-Simons actions $I_{\text{CS}}[A^+]-I_{\text{CS}}[A^-]$. We adopt the boundary terms $I_{\text{bdy}}^+[A^+]$ for $A^+$ and $I^-_{\text{bdy}}[A^-]$ for $A^-$, and fix the components $A^+_z$ and $A^-_{\bar{z}}$ as the corresponding form in \eqref{bdycomponents}. The remaining components $A^+_{\bar{z}}, A^-_{z}$ are automatically determined from the condition that $A^+,A^-$ are anti-hermitian.

After all, the total action we consider is 
\begin{align}\label{totalactionB}
    I_{\text{G}}&=I_{\text{CS}}[A^+]-I_{\text{CS}}[A^-]+\frac{k}{4\pi}\int_{\theta=\pi/2}\Tr[A^+_zA^+_{\bar{z}}+A^-_zA^-_{\bar{z}}].
\end{align}
Each on-shell action for the solution \eqref{FSconnection} is evaluated as 
\begin{align}
    &I_{\text{CS}}^{\text{on-shell}}[A^+]=-I_{\text{CS}}^{\text{on-shell}}[A^-]=-\frac{\pi }{8G_N},\\
    &\int_{\theta=\pi/2}\Tr[A^+_zA^+_{\bar{z}}+A^-_zA^-_{\bar{z}}]=0
\end{align}
Thus we have 
\begin{align}
    I_{\text{G}}^{\text{on-shell}}[B^3]=-\frac{\pi }{4G_N},
\end{align}
the twice of which is equal to the on-shell action for the whole $S^3$ spacetime. 

The Fubini-Study metric of $S^2$ is transformed to the ordinary polar coordinate 
\begin{align}
    ds_{S^2}^2=d\psi^2+\sin^2\psi d\varphi^2
\end{align}
by a coordinate transformation $z=e^{i\varphi}/\tan(\psi/2)$. In that coordinate, the on-shell action becomes 
\begin{align}
    \frac{ik}{2\pi} \int_{\theta=\frac{\pi}{2}}d\psi d\varphi \frac{\cos(\psi/2)}{2\sin^3(\psi/2)}\Tr[A^+_zA^+_{\bar{z}}+A^-_zA^-_{\bar{z}}]
\end{align}
By replacing with $\varphi\to r_0\varphi$, we can make the conical defect geometry. Since the integrand in the above on-shell action does not depend on $\varphi$, the scaling $\varphi\to r_0\varphi$ just gives the ovarall factor $r_0$ coming from the measure. Thus the on-shell action on such a geometry is 
\begin{align}
    I_{\text{G}}^{\text{on-shell}}[\text{BH}_{r_0}]=-\frac{\pi r_0}{4G_N}
\end{align}
Pictorially this geometry is topologically equivalent to $B^3$ but along a conical defect $\phi=0$ approaching to the boundary. 

\subsection{Relation to WZW model}
We would like to discuss the reduction to WZW model. Since the boundary of the space we are considering is not an asymptotic boundary, we adopt a different method which is described in \cite{Arcioni:2002vv}. 

The background solutions \eqref{FSconnection} can be decomposed as 
\begin{align}
    A^+=b^{-1}db+b^{-1}ab,\qquad A^-=bdb^{-1}+bab^{-1},
\end{align}
by using same 1-form $a$ and $b=e^{\theta J_2}$, and $a$ can be written as the pure gauge 
\begin{align}
    a_z=h^{-1}\partial_zh,\qquad a_{\bar{z}}=h^{-1}\partial_{\bar{z}} h,
\end{align}
where 
\begin{align}
    h=\frac{1}{\sqrt{1+|z|^2}}
    \begin{pmatrix}
        1 & -z\\
        \bar{z} & 1
    \end{pmatrix}.
\end{align}
Thus $A^{\pm}$ themselves can also be expressed as 
\begin{align}
    A^{\pm}=(G^{\pm})^{-1}dG^{\pm},
\end{align}
where 
\begin{align}
    G^+=hb,\qquad G^-=hb^{-1}.
\end{align}
Substituting the expressions to Chern-Simons gravity action \eqref{totalactionB} leads to \cite{Arcioni:2002vv}
\begin{align}
    I_{\text{G}}[G_+^{-1}dG_+,G_-^{-1}dG_-]=I_{\text{WZW}}[G_+G_-^{-1}],
\end{align}
where $I_{\text{WZW}}[g]$ denotes the usual non-chiral WZW action. We can derive the relation by using the Polyakov-Wiegman identity.

\bibliographystyle{JHEP}
\bibliography{Refs}

\providecommand{\href}[2]{#2}\begingroup\raggedright\begin{thebibliography}{10}

\bibitem{Maldacena:1997re}
J.M.~Maldacena, \emph{{The Large N limit of superconformal field theories and
  supergravity}}, \href{https://doi.org/10.4310/ATMP.1998.v2.n2.a1}{\emph{Adv.
  Theor. Math. Phys.} {\bfseries 2} (1998) 231}
  [\href{https://arxiv.org/abs/hep-th/9711200}{{\ttfamily hep-th/9711200}}].

\bibitem{Gubser:1998bc}
S.S.~Gubser, I.R.~Klebanov and A.M.~Polyakov, \emph{{Gauge theory correlators
  from noncritical string theory}},
  \href{https://doi.org/10.1016/S0370-2693(98)00377-3}{\emph{Phys. Lett. B}
  {\bfseries 428} (1998) 105}
  [\href{https://arxiv.org/abs/hep-th/9802109}{{\ttfamily hep-th/9802109}}].

\bibitem{Witten:1998qj}
E.~Witten, \emph{{Anti-de Sitter space and holography}},
  \href{https://doi.org/10.4310/ATMP.1998.v2.n2.a2}{\emph{Adv. Theor. Math.
  Phys.} {\bfseries 2} (1998) 253}
  [\href{https://arxiv.org/abs/hep-th/9802150}{{\ttfamily hep-th/9802150}}].

\bibitem{Strominger:2001pn}
A.~Strominger, \emph{{The dS / CFT correspondence}},
  \href{https://doi.org/10.1088/1126-6708/2001/10/034}{\emph{JHEP} {\bfseries
  10} (2001) 034} [\href{https://arxiv.org/abs/hep-th/0106113}{{\ttfamily
  hep-th/0106113}}].

\bibitem{Maldacena:2002vr}
J.M.~Maldacena, \emph{{Non-Gaussian features of primordial fluctuations in
  single field inflationary models}},
  \href{https://doi.org/10.1088/1126-6708/2003/05/013}{\emph{JHEP} {\bfseries
  05} (2003) 013} [\href{https://arxiv.org/abs/astro-ph/0210603}{{\ttfamily
  astro-ph/0210603}}].

\bibitem{Witten:2001kn}
E.~Witten, \emph{{Quantum gravity in de Sitter space}},  in \emph{{Strings
  2001: International Conference}}, 6, 2001
  [\href{https://arxiv.org/abs/hep-th/0106109}{{\ttfamily hep-th/0106109}}].

\bibitem{Pasterski:2016qvg}
S.~Pasterski, S.-H.~Shao and A.~Strominger, \emph{{Flat Space Amplitudes and
  Conformal Symmetry of the Celestial Sphere}},
  \href{https://doi.org/10.1103/PhysRevD.96.065026}{\emph{Phys. Rev. D}
  {\bfseries 96} (2017) 065026}
  [\href{https://arxiv.org/abs/1701.00049}{{\ttfamily 1701.00049}}].

\bibitem{Raclariu:2021zjz}
A.-M.~Raclariu, \emph{{Lectures on Celestial Holography}},
  \href{https://arxiv.org/abs/2107.02075}{{\ttfamily 2107.02075}}.

\bibitem{Pasterski:2021rjz}
S.~Pasterski, \emph{{Lectures on celestial amplitudes}},
  \href{https://doi.org/10.1140/epjc/s10052-021-09846-7}{\emph{Eur. Phys. J. C}
  {\bfseries 81} (2021) 1062}
  [\href{https://arxiv.org/abs/2108.04801}{{\ttfamily 2108.04801}}].

\bibitem{Kawamoto:2023wzj}
T.~Kawamoto, S.-M.~Ruan and T.~Takayanagia, \emph{{Gluing AdS/CFT}},
  \href{https://doi.org/10.1007/JHEP07(2023)080}{\emph{JHEP} {\bfseries 07}
  (2023) 080} [\href{https://arxiv.org/abs/2303.01247}{{\ttfamily
  2303.01247}}].

\bibitem{Hikida:2021ese}
Y.~Hikida, T.~Nishioka, T.~Takayanagi and Y.~Taki, \emph{{Holography in de
  Sitter Space via Chern-Simons Gauge Theory}},
  \href{https://doi.org/10.1103/PhysRevLett.129.041601}{\emph{Phys. Rev. Lett.}
  {\bfseries 129} (2022) 041601}
  [\href{https://arxiv.org/abs/2110.03197}{{\ttfamily 2110.03197}}].

\bibitem{Hikida:2022ltr}
Y.~Hikida, T.~Nishioka, T.~Takayanagi and Y.~Taki, \emph{{CFT duals of
  three-dimensional de Sitter gravity}},
  \href{https://doi.org/10.1007/JHEP05(2022)129}{\emph{JHEP} {\bfseries 05}
  (2022) 129} [\href{https://arxiv.org/abs/2203.02852}{{\ttfamily
  2203.02852}}].

\bibitem{Hartle:1983ai}
J.B.~Hartle and S.W.~Hawking, \emph{{Wave Function of the Universe}},
  \href{https://doi.org/10.1103/PhysRevD.28.2960}{\emph{Phys. Rev. D}
  {\bfseries 28} (1983) 2960}.

\bibitem{Almheiri:2014cka}
A.~Almheiri and J.~Polchinski, \emph{{Models of AdS$_{2}$ backreaction and
  holography}}, \href{https://doi.org/10.1007/JHEP11(2015)014}{\emph{JHEP}
  {\bfseries 11} (2015) 014} [\href{https://arxiv.org/abs/1402.6334}{{\ttfamily
  1402.6334}}].

\bibitem{Jensen:2016pah}
K.~Jensen, \emph{{Chaos in AdS$_2$ Holography}},
  \href{https://doi.org/10.1103/PhysRevLett.117.111601}{\emph{Phys. Rev. Lett.}
  {\bfseries 117} (2016) 111601}
  [\href{https://arxiv.org/abs/1605.06098}{{\ttfamily 1605.06098}}].

\bibitem{Maldacena:2016upp}
J.~Maldacena, D.~Stanford and Z.~Yang, \emph{{Conformal symmetry and its
  breaking in two dimensional Nearly Anti-de-Sitter space}},
  \href{https://doi.org/10.1093/ptep/ptw124}{\emph{PTEP} {\bfseries 2016}
  (2016) 12C104} [\href{https://arxiv.org/abs/1606.01857}{{\ttfamily
  1606.01857}}].

\bibitem{Engelsoy:2016xyb}
J.~Engels\"oy, T.G.~Mertens and H.~Verlinde, \emph{{An investigation of
  AdS$_{2}$ backreaction and holography}},
  \href{https://doi.org/10.1007/JHEP07(2016)139}{\emph{JHEP} {\bfseries 07}
  (2016) 139} [\href{https://arxiv.org/abs/1606.03438}{{\ttfamily
  1606.03438}}].

\bibitem{Ghosh:2019rcj}
A.~Ghosh, H.~Maxfield and G.J.~Turiaci, \emph{{A universal Schwarzian sector in
  two-dimensional conformal field theories}},
  \href{https://doi.org/10.1007/JHEP05(2020)104}{\emph{JHEP} {\bfseries 05}
  (2020) 104} [\href{https://arxiv.org/abs/1912.07654}{{\ttfamily
  1912.07654}}].

\bibitem{Jackiw:1984je}
R.~Jackiw, \emph{{Lower Dimensional Gravity}},
  \href{https://doi.org/10.1016/0550-3213(85)90448-1}{\emph{Nucl. Phys.}
  {\bfseries B252} (1985) 343}.

\bibitem{Teitelboim:1983ux}
C.~Teitelboim, \emph{{Gravitation and Hamiltonian Structure in Two Space-Time
  Dimensions}}, \href{https://doi.org/10.1016/0370-2693(83)90012-6}{\emph{Phys.
  Lett.} {\bfseries 126B} (1983) 41}.

\bibitem{Maldacena:2016hyu}
J.~Maldacena and D.~Stanford, \emph{{Remarks on the Sachdev-Ye-Kitaev model}},
  \href{https://doi.org/10.1103/PhysRevD.94.106002}{\emph{Phys. Rev. D}
  {\bfseries 94} (2016) 106002}
  [\href{https://arxiv.org/abs/1604.07818}{{\ttfamily 1604.07818}}].

\bibitem{Mertens:2017mtv}
T.G.~Mertens, G.J.~Turiaci and H.L.~Verlinde, \emph{{Solving the Schwarzian via
  the Conformal Bootstrap}},
  \href{https://doi.org/10.1007/JHEP08(2017)136}{\emph{JHEP} {\bfseries 08}
  (2017) 136} [\href{https://arxiv.org/abs/1705.08408}{{\ttfamily
  1705.08408}}].

\bibitem{Mertens:2018fds}
T.G.~Mertens, \emph{{The Schwarzian theory \textemdash{} origins}},
  \href{https://doi.org/10.1007/JHEP05(2018)036}{\emph{JHEP} {\bfseries 05}
  (2018) 036} [\href{https://arxiv.org/abs/1801.09605}{{\ttfamily
  1801.09605}}].

\bibitem{Maldacena:1998ih}
J.M.~Maldacena and A.~Strominger, \emph{{Statistical entropy of de Sitter
  space}}, \href{https://doi.org/10.1088/1126-6708/1998/02/014}{\emph{JHEP}
  {\bfseries 02} (1998) 014}
  [\href{https://arxiv.org/abs/gr-qc/9801096}{{\ttfamily gr-qc/9801096}}].

\bibitem{Park:1998qk}
M.-I.~Park, \emph{{Statistical entropy of three-dimensional Kerr-de Sitter
  space}}, \href{https://doi.org/10.1016/S0370-2693(98)01119-8}{\emph{Phys.
  Lett. B} {\bfseries 440} (1998) 275}
  [\href{https://arxiv.org/abs/hep-th/9806119}{{\ttfamily hep-th/9806119}}].

\bibitem{Park:1998yw}
M.-I.~Park, \emph{{Symmetry algebras in Chern-Simons theories with boundary:
  Canonical approach}},
  \href{https://doi.org/10.1016/S0550-3213(99)00031-0}{\emph{Nucl. Phys. B}
  {\bfseries 544} (1999) 377}
  [\href{https://arxiv.org/abs/hep-th/9811033}{{\ttfamily hep-th/9811033}}].

\bibitem{Anninos:2011ui}
D.~Anninos, T.~Hartman and A.~Strominger, \emph{{Higher Spin Realization of the
  dS/CFT Correspondence}},
  \href{https://doi.org/10.1088/1361-6382/34/1/015009}{\emph{Class. Quant.
  Grav.} {\bfseries 34} (2017) 015009}
  [\href{https://arxiv.org/abs/1108.5735}{{\ttfamily 1108.5735}}].

\bibitem{Klebanov:2002ja}
I.R.~Klebanov and A.M.~Polyakov, \emph{{AdS dual of the critical O(N) vector
  model}}, \href{https://doi.org/10.1016/S0370-2693(02)02980-5}{\emph{Phys.
  Lett. B} {\bfseries 550} (2002) 213}
  [\href{https://arxiv.org/abs/hep-th/0210114}{{\ttfamily hep-th/0210114}}].

\bibitem{Gaberdiel:2010pz}
M.R.~Gaberdiel and R.~Gopakumar, \emph{{An AdS$_{3}$ Dual for Minimal Model
  CFTs}}, \href{https://doi.org/10.1103/PhysRevD.83.066007}{\emph{Phys. Rev. D}
  {\bfseries 83} (2011) 066007}
  [\href{https://arxiv.org/abs/1011.2986}{{\ttfamily 1011.2986}}].

\bibitem{Gaberdiel:2012ku}
M.R.~Gaberdiel and R.~Gopakumar, \emph{{Triality in Minimal Model Holography}},
  \href{https://doi.org/10.1007/JHEP07(2012)127}{\emph{JHEP} {\bfseries 07}
  (2012) 127} [\href{https://arxiv.org/abs/1205.2472}{{\ttfamily 1205.2472}}].

\bibitem{Gaberdiel:2012uj}
M.R.~Gaberdiel and R.~Gopakumar, \emph{{Minimal Model Holography}},
  \href{https://doi.org/10.1088/1751-8113/46/21/214002}{\emph{J. Phys. A}
  {\bfseries 46} (2013) 214002}
  [\href{https://arxiv.org/abs/1207.6697}{{\ttfamily 1207.6697}}].

\bibitem{Perlmutter:2012ds}
E.~Perlmutter, T.~Prochazka and J.~Raeymaekers, \emph{{The semiclassical limit
  of $W_N$ CFTs and Vasiliev theory}},
  \href{https://doi.org/10.1007/JHEP05(2013)007}{\emph{JHEP} {\bfseries 05}
  (2013) 007} [\href{https://arxiv.org/abs/1210.8452}{{\ttfamily 1210.8452}}].

\bibitem{Witten:2021nzp}
E.~Witten, \emph{{A Note On Complex Spacetime Metrics}},
  \href{https://arxiv.org/abs/2111.06514}{{\ttfamily 2111.06514}}.

\bibitem{Chen:2023prz}
H.-Y.~Chen, Y.~Hikida, Y.~Taki and T.~Uetoko, \emph{{Complex saddles of
  three-dimensional de Sitter gravity via holography}},
  \href{https://doi.org/10.1103/PhysRevD.107.L101902}{\emph{Phys. Rev. D}
  {\bfseries 107} (2023) L101902}
  [\href{https://arxiv.org/abs/2302.09219}{{\ttfamily 2302.09219}}].

\bibitem{Chen:2023sry}
H.-Y.~Chen, Y.~Hikida, Y.~Taki and T.~Uetoko, \emph{{Complex saddles of
  Chern-Simons gravity and dS$_3$/CFT$_2$ correspondence}},
  \href{https://arxiv.org/abs/2306.03330}{{\ttfamily 2306.03330}}.

\bibitem{Gibbons:1976ue}
G.W.~Gibbons and S.W.~Hawking, \emph{{Action Integrals and Partition Functions
  in Quantum Gravity}},
  \href{https://doi.org/10.1103/PhysRevD.15.2752}{\emph{Phys. Rev. D}
  {\bfseries 15} (1977) 2752}.

\bibitem{Gibbons:1977mu}
G.W.~Gibbons and S.W.~Hawking, \emph{{Cosmological Event Horizons,
  Thermodynamics, and Particle Creation}},
  \href{https://doi.org/10.1103/PhysRevD.15.2738}{\emph{Phys. Rev. D}
  {\bfseries 15} (1977) 2738}.

\bibitem{Maldacena:2019cbz}
J.~Maldacena, G.J.~Turiaci and Z.~Yang, \emph{{Two dimensional Nearly de Sitter
  gravity}}, \href{https://doi.org/10.1007/JHEP01(2021)139}{\emph{JHEP}
  {\bfseries 01} (2021) 139}
  [\href{https://arxiv.org/abs/1904.01911}{{\ttfamily 1904.01911}}].

\bibitem{Goel:2020yxl}
A.~Goel, L.V.~Iliesiu, J.~Kruthoff and Z.~Yang, \emph{{Classifying boundary
  conditions in JT gravity: from energy-branes to $\alpha$-branes}},
  \href{https://doi.org/10.1007/JHEP04(2021)069}{\emph{JHEP} {\bfseries 04}
  (2021) 069} [\href{https://arxiv.org/abs/2010.12592}{{\ttfamily
  2010.12592}}].

\bibitem{Susskind:2021omt}
L.~Susskind, \emph{{De Sitter Holography: Fluctuations, Anomalous Symmetry, and
  Wormholes}}, \href{https://doi.org/10.3390/universe7120464}{\emph{Universe}
  {\bfseries 7} (2021) 464} [\href{https://arxiv.org/abs/2106.03964}{{\ttfamily
  2106.03964}}].

\bibitem{Susskind:2021dfc}
L.~Susskind, \emph{{Black Holes Hint towards De Sitter Matrix Theory}},
  \href{https://doi.org/10.3390/universe9080368}{\emph{Universe} {\bfseries 9}
  (2023) 368} [\href{https://arxiv.org/abs/2109.01322}{{\ttfamily
  2109.01322}}].

\bibitem{Kawamoto:2023nki}
T.~Kawamoto, S.-M.~Ruan, Y.-k.~Suzuki and T.~Takayanagi, \emph{{A Half de
  Sitter Holography}},  \href{https://arxiv.org/abs/2306.07575}{{\ttfamily
  2306.07575}}.

\bibitem{Mertens:2019tcm}
T.G.~Mertens and G.J.~Turiaci, \emph{{Defects in Jackiw-Teitelboim Quantum
  Gravity}}, \href{https://doi.org/10.1007/JHEP08(2019)127}{\emph{JHEP}
  {\bfseries 08} (2019) 127}
  [\href{https://arxiv.org/abs/1904.05228}{{\ttfamily 1904.05228}}].

\bibitem{Maxfield:2020ale}
H.~Maxfield and G.J.~Turiaci, \emph{{The path integral of 3D gravity near
  extremality; or, JT gravity with defects as a matrix integral}},
  \href{https://doi.org/10.1007/JHEP01(2021)118}{\emph{JHEP} {\bfseries 01}
  (2021) 118} [\href{https://arxiv.org/abs/2006.11317}{{\ttfamily
  2006.11317}}].

\bibitem{Witten:2020wvy}
E.~Witten, \emph{{Matrix Models and Deformations of JT Gravity}},
  \href{https://doi.org/10.1098/rspa.2020.0582}{\emph{Proc. Roy. Soc. Lond. A}
  {\bfseries 476} (2020) 20200582}
  [\href{https://arxiv.org/abs/2006.13414}{{\ttfamily 2006.13414}}].

\bibitem{Mefford:2020vde}
E.~Mefford and K.~Suzuki, \emph{{Jackiw-Teitelboim quantum gravity with defects
  and the Aharonov-Bohm effect}},
  \href{https://doi.org/10.1007/JHEP05(2021)026}{\emph{JHEP} {\bfseries 05}
  (2021) 026} [\href{https://arxiv.org/abs/2011.04695}{{\ttfamily
  2011.04695}}].

\bibitem{Achucarro:1986uwr}
A.~Achucarro and P.K.~Townsend, \emph{{A Chern-Simons Action for
  Three-Dimensional anti-De Sitter Supergravity Theories}},
  \href{https://doi.org/10.1016/0370-2693(86)90140-1}{\emph{Phys. Lett. B}
  {\bfseries 180} (1986) 89}.

\bibitem{Witten:1988hc}
E.~Witten, \emph{{(2+1)-Dimensional Gravity as an Exactly Soluble System}},
  \href{https://doi.org/10.1016/0550-3213(88)90143-5}{\emph{Nucl. Phys. B}
  {\bfseries 311} (1988) 46}.

\bibitem{Coussaert:1995zp}
O.~Coussaert, M.~Henneaux and P.~van Driel, \emph{{The Asymptotic dynamics of
  three-dimensional Einstein gravity with a negative cosmological constant}},
  \href{https://doi.org/10.1088/0264-9381/12/12/012}{\emph{Class. Quant. Grav.}
  {\bfseries 12} (1995) 2961}
  [\href{https://arxiv.org/abs/gr-qc/9506019}{{\ttfamily gr-qc/9506019}}].

\bibitem{Donnay:2016iyk}
L.~Donnay, \emph{{Asymptotic dynamics of three-dimensional gravity}},
  \href{https://doi.org/10.22323/1.271.0001}{\emph{PoS} {\bfseries Modave2015}
  (2016) 001} [\href{https://arxiv.org/abs/1602.09021}{{\ttfamily
  1602.09021}}].

\bibitem{Achucarro:1992mb}
A.~Achucarro, \emph{{Lineal gravity from planar gravity}},
  \href{https://doi.org/10.1103/PhysRevLett.70.1037}{\emph{Phys. Rev. Lett.}
  {\bfseries 70} (1993) 1037}
  [\href{https://arxiv.org/abs/hep-th/9207108}{{\ttfamily hep-th/9207108}}].

\bibitem{Isler:1989hq}
K.~Isler and C.A.~Trugenberger, \emph{{A Gauge Theory of Two-dimensional
  Quantum Gravity}},
  \href{https://doi.org/10.1103/PhysRevLett.63.834}{\emph{Phys. Rev. Lett.}
  {\bfseries 63} (1989) 834}.

\bibitem{Chamseddine:1989yz}
A.H.~Chamseddine and D.~Wyler, \emph{{Gauge Theory of Topological Gravity in
  (1+1)-Dimensions}},
  \href{https://doi.org/10.1016/0370-2693(89)90528-5}{\emph{Phys. Lett. B}
  {\bfseries 228} (1989) 75}.

\bibitem{Iliesiu:2019xuh}
L.V.~Iliesiu, S.S.~Pufu, H.~Verlinde and Y.~Wang, \emph{{An exact quantization
  of Jackiw-Teitelboim gravity}},
  \href{https://doi.org/10.1007/JHEP11(2019)091}{\emph{JHEP} {\bfseries 11}
  (2019) 091} [\href{https://arxiv.org/abs/1905.02726}{{\ttfamily
  1905.02726}}].

\bibitem{Vasiliev:1995sv}
M.A.~Vasiliev, \emph{{Higher spin gauge interactions for matter fields in
  two-dimensions}},
  \href{https://doi.org/10.1016/0370-2693(95)01122-7}{\emph{Phys. Lett. B}
  {\bfseries 363} (1995) 51}
  [\href{https://arxiv.org/abs/hep-th/9511063}{{\ttfamily hep-th/9511063}}].

\bibitem{Alekseev:1988ce}
A.~Alekseev and S.L.~Shatashvili, \emph{{Path Integral Quantization of the
  Coadjoint Orbits of the Virasoro Group and 2D Gravity}},
  \href{https://doi.org/10.1016/0550-3213(89)90130-2}{\emph{Nucl. Phys. B}
  {\bfseries 323} (1989) 719}.

\bibitem{Castro:2011iw}
A.~Castro, R.~Gopakumar, M.~Gutperle and J.~Raeymaekers, \emph{{Conical Defects
  in Higher Spin Theories}},
  \href{https://doi.org/10.1007/JHEP02(2012)096}{\emph{JHEP} {\bfseries 02}
  (2012) 096} [\href{https://arxiv.org/abs/1111.3381}{{\ttfamily 1111.3381}}].

\bibitem{Cotler:2018zff}
J.~Cotler and K.~Jensen, \emph{{A theory of reparameterizations for AdS$_3$
  gravity}}, \href{https://doi.org/10.1007/JHEP02(2019)079}{\emph{JHEP}
  {\bfseries 02} (2019) 079}
  [\href{https://arxiv.org/abs/1808.03263}{{\ttfamily 1808.03263}}].

\bibitem{Cotler:2019nbi}
J.~Cotler, K.~Jensen and A.~Maloney, \emph{{Low-dimensional de Sitter quantum
  gravity}}, \href{https://doi.org/10.1007/JHEP06(2020)048}{\emph{JHEP}
  {\bfseries 06} (2020) 048}
  [\href{https://arxiv.org/abs/1905.03780}{{\ttfamily 1905.03780}}].

\bibitem{Creutzig:2021ykz}
T.~Creutzig and Y.~Hikida, \emph{{Correlator correspondences for
  Gaiotto-Rap\v{c}\'ak dualities and first order formulation of coset models}},
  \href{https://doi.org/10.1007/JHEP12(2021)144}{\emph{JHEP} {\bfseries 12}
  (2021) 144} [\href{https://arxiv.org/abs/2109.03403}{{\ttfamily
  2109.03403}}].

\bibitem{Witten:1988hf}
E.~Witten, \emph{{Quantum Field Theory and the Jones Polynomial}},
  \href{https://doi.org/10.1007/BF01217730}{\emph{Commun. Math. Phys.}
  {\bfseries 121} (1989) 351}.

\bibitem{Moore:1989yh}
G.W.~Moore and N.~Seiberg, \emph{{Taming the Conformal Zoo}},
  \href{https://doi.org/10.1016/0370-2693(89)90897-6}{\emph{Phys. Lett. B}
  {\bfseries 220} (1989) 422}.

\bibitem{Balog:1997zz}
J.~Balog, L.~Feher and L.~Palla, \emph{{Coadjoint orbits of the Virasoro
  algebra and the global Liouville equation}},
  \href{https://doi.org/10.1142/S0217751X98000147}{\emph{Int. J. Mod. Phys. A}
  {\bfseries 13} (1998) 315}
  [\href{https://arxiv.org/abs/hep-th/9703045}{{\ttfamily hep-th/9703045}}].

\bibitem{Forgacs:1989ac}
P.~Forgacs, A.~Wipf, J.~Balog, L.~Feher and L.~O'Raifeartaigh, \emph{{Liouville
  and Toda Theories as Conformally Reduced WZNW Theories}},
  \href{https://doi.org/10.1016/S0370-2693(89)80025-5}{\emph{Phys. Lett. B}
  {\bfseries 227} (1989) 214}.

\bibitem{Suzuki:2023}
K.~Suzuki and Y.~Taki, \emph{{Correlation functions in $S^2$ JT gravity and
  complex Liouville quamtum mechanics}}, {\emph{in progress \hspace*{-7pt}} }.

\bibitem{Maldacena:2015waa}
J.~Maldacena, S.H.~Shenker and D.~Stanford, \emph{{A bound on chaos}},
  \href{https://doi.org/10.1007/JHEP08(2016)106}{\emph{JHEP} {\bfseries 08}
  (2016) 106} [\href{https://arxiv.org/abs/1503.01409}{{\ttfamily
  1503.01409}}].

\bibitem{DiVecchia:1984nyg}
P.~Di~Vecchia, V.G.~Knizhnik, J.L.~Petersen and P.~Rossi, \emph{{A
  Supersymmetric Wess-Zumino Lagrangian in Two-Dimensions}},
  \href{https://doi.org/10.1016/0550-3213(85)90554-1}{\emph{Nucl. Phys. B}
  {\bfseries 253} (1985) 701}.

\bibitem{Kac:1985wdn}
V.G.~Kac and I.T.~Todorov, \emph{{SUPERCONFORMAL CURRENT ALGEBRAS AND THEIR
  UNITARY REPRESENTATIONS}},
  \href{https://doi.org/10.1007/BF01229384}{\emph{Commun. Math. Phys.}
  {\bfseries 102} (1985) 337}.

\bibitem{Fuchs:1986ew}
J.~Fuchs, \emph{{Superconformal Ward Identities and the {WZW} Model}},
  \href{https://doi.org/10.1016/0550-3213(87)90450-0}{\emph{Nucl. Phys. B}
  {\bfseries 286} (1987) 455}.

\bibitem{Arcioni:2002vv}
G.~Arcioni, M.~Blau and M.~O'Loughlin, \emph{{On the boundary dynamics of
  Chern-Simons gravity}},
  \href{https://doi.org/10.1088/1126-6708/2003/01/067}{\emph{JHEP} {\bfseries
  01} (2003) 067} [\href{https://arxiv.org/abs/hep-th/0210089}{{\ttfamily
  hep-th/0210089}}].

\end{thebibliography}\endgroup


\end{document}